\documentclass[unnumsec,webpdf,modern,large]{oup-authoring-template-blank}%

\usepackage[T1]{fontenc}
\usepackage[utf8]{inputenc}

\usepackage{listings}

\usepackage{alphalph}
\newcounter{QuoteCounter}
\newcommand{\quoteletter}{\stepcounter{QuoteCounter}\AlphAlph{\value{QuoteCounter}}}

\usepackage{todonotes}%
\let\OldTodo\todo
\renewcommand{\todo}{\OldTodo[inline]}%
\newcommand{\todolater}[1]{}%

\theoremstyle{thmstyleone}%
\theoremstyle{thmstyletwo}%
\theoremstyle{thmstylethree}%

\begin{document}

\firstpage{1}

\title[Stand-Alone Complex or Vibercrime?]{Stand-Alone Complex or Vibercrime? Exploring the adoption and innovation of GenAI tools, coding assistants, and agents within cybercrime ecosystems}

 \author[1,$\ast$]{Jack Hughes\ORCID{0000-0002-0730-1055}}
 \author[2,1]{Ben Collier\ORCID{0000-0002-9207-3068}}
 \author[3,1]{Daniel R. Thomas\ORCID{0000-0001-8936-0683}}

 \authormark{Jack Hughes et al.}

 \address[1]{\orgdiv{Cambridge Cybercrime Centre, Department of Computer Science \& Technology}, \orgname{University of Cambridge}, \orgaddress{\postcode{CB3 0FD}, \country{UK}}}
  \address[2]{\orgdiv{School of Social and Political Science}, \orgname{University of Edinburgh}, \orgaddress{\postcode{EH8 9LD}, \country{UK}}}
 \address[3]{\orgdiv{Computer \& Information Sciences}, \orgname{University of Strathclyde, Glasgow}, \orgaddress{\postcode{G1 1XH}, \country{UK}}}

 \corresp[$\ast$]{Corresponding author. \href{email:firstname.lastname@cl.cam.ac.uk}{firstname.lastname@cl.cam.ac.uk}}

\received{Date}{0}{Year}
\revised{Date}{0}{Year}
\accepted{Date}{0}{Year}

\abstract{Existential risk scenarios relating to Generative Artificial Intelligence often involve advanced systems or agentic models breaking loose and using hacking tools to gain control over critical infrastructure. We argue that the real threats posed by generative AI for cybercrime are rather different. We apply innovation theory and evolutionary economics -- treating cybercrime as an ecosystem of small- and medium-scale tech start-ups, coining two novel terms that bound the upper and lower cases for disruption. At the high end, we propose the Stand-Alone Complex, in which cybercrime-gang-in-a-box solutions enable individual actors to further automate cybercrime-as-a-service arrangements. At the low end, we suggest the phenomenon of Vibercrime, in which `vibe coding' chatbot assistants lower the barrier to entry, but do not fundamentally reshape the economic structures of cybercrime. We analyse early empirical data from a variety of large-scale digital sources from the cybercrime underground, and find that the reality is prosaic -- AI is seeing some early adoption in existing large-scale, low-profit passive income schemes and forms of fraud but there is little evidence that it is so far giving rise to widespread disruption in technical aspects of cybercrime. It is also not being widely used to overcome skill barriers or as innovative disruptor for cybercrime-specific coding domains (which already rely heavily on old, pre-made resources, scripts, and exploits). Instead, it is replacing existing means of code pasting, error checking, and cheatsheet consultation, mostly for generic aspects of software development involved in cybercrime -- and largely for already skilled actors, with low-skill actors finding little utility in vibe coding tools compared to pre-made scripts. The role of jailbroken LLMs (Dark AI) as hacking instructors is also overstated, given the prominence of subculture and social learning in initiation -- new users value the social connections and community identity involved in learning hacking and cybercrime skills as much as the knowledge itself. Disruptive adoption is mainly in the domain of logistics and automation, especially in bot farming and the production of approach emails and social frauds.  Our initial results, therefore, suggest an intensification and scaling of the dynamics of cybercrime-as-a-service rather than a rupture.}


\maketitle

\section{Introduction}
AI risk -- and the so-called `existential' risks of AI -- have become central objects of policy and practitioner scrutiny since the release of ChatGPT in 2022 \cite{analytica_policy_2023,bareis_ai_2025}. Where technological innovation produces novel harms, or the negative externalities -- climate, environmental, health, social, or political -- of established technologies and infrastructures become increasingly apparent, it is vital that industry and government have the tools to respond. Whether through policy, regulation, legislation, or interventions in the design of technologies, this often invokes the \textit{precautionary principle} (and the related practices of Responsible Research and Innovation) -- namely, that regulators and developers of technology should act to anticipate harms \textit{before} they become widespread, rather than solely in reaction to mass-scale illness, pollution, or injury \cite{florin_risk_2022}. Bringing this principle into the management of risks posed by innovations in artificial intelligence technologies, however, is proving problematic -- not least in the impacts these technologies may have on cybercrime \cite{kasirzadeh_two_2025,vredenburgh_ai_2025}.

Wildly diverging views of the potential trajectories of the technology and the potential harms that might result from different pathways are being bound up with the hype that has become a core functional component of the tech ecosystem \cite{bareis_ai_2025,westerstrand_talking_2024,gebru_tescreal_2024}. In particular, the (entirely hypothetical) spectacle of Artificial General Intelligence -- in which AI systems develop exponential capacities (either decisively or accumulatively) \cite{kasirzadeh_two_2025} to improve themselves and establish control over the world -- is a core risk case for many of the CEOs, policymakers, and politicians shaping the current industry \footnote{See for example, a January 2026 paper in the UK House of Lords Library on `Superintelligent AI: Should its development be stopped?'~\cite{evennett_superintelligent_2026} and the UK Government Office for Science's AI 2030 Scenarios~\cite{noauthor_ai_2024}}. This `criti-hype'~\cite{vinsel_youre_2021} serves both to inflate the apparent power and importance of the technology for the purposes of attracting venture capital funding and investment, and encourages strategic regulatory lock-in, meaning that only large and established corporate players are able to compete \cite{gebru_tescreal_2024}.

The tools, technologies, and methods of the cybercrime ecosystem play a major role in many of the `existential risk' scenarios of artificial intelligence trajectories~\cite{kasirzadeh_two_2025}. Most of the critical systems and national infrastructure on which countries rely are dependent on, or connected to, digital infrastructure. Many existential risk AGI scenarios, therefore, involve `breaking containment', connecting to the wider Internet and taking control of physical systems, financial, and computational resources. These scenarios seem to be science fiction -- while a few pre-release models in testing have been shown to escape sandboxes\cite{openai_huggingface_2026}, available LLMs have not yet shown this capability -- however they are nonetheless exerting a strong shaping force on regulation \cite{kasirzadeh_singularity_2025}.

More prosaic impacts on cybercrime, however, are not inconceivable. For example, something approaching a step-change in the cybercrime landscape could result from increasing automation within the current cybercrime ecosystem -- with existing cybercrime-as-a-service business models being packaged up into `crime-gang-in-a-box' infrastructure that could be sold as a passive income generator on cybercrime forums. Indeed, in areas of cybercrime where large-scale automation already competes with the platforms' own algorithms, such as SEO fraud, serious disruption can be readily observed \cite{burton_ai_2025}.

Less florid, though still speculative, potential risk scenarios involve the adoption of GenAI tools leading to progressive smaller-scale improvements in the technical and organisational capabilities of the existing cybercrime underground \cite{kasirzadeh_two_2025}. The possibility of generative AI tools leading to an uplift in the ability of states, organised crime gangs, small-scale cybercriminals and neophytes\footnote{A neophyte is someone who has recently joined a subject or community.} (or `skids') to disrupt the foundations of digital society through cybercrime is another high-risk scenario being considered by policymakers and engineers.

We present here one of the first attempts at a mixed-methods empirical study of early patterns of GenAI adoption in the cybercrime underground. We first discuss the criminological literature on cybercrime, and on innovation theory and evolutionary economics, and then set out a theoretical framework for making sense of GenAI adoption in the cybercrime underground, further discussing existing research to date on this topic. We then empirically explore how these new technologies -- particularly implementations in legitimate industry -- are actually being adopted in the cybercrime ecosystem. Drawing on the CrimeBB dataset~\cite{pastrana_crimebb_2018}, comprising more than 100 million forum posts and chat channel discussions from the cybercrime underground over more than 15 years, we are able to trace a wider set of historic and present innovation trajectories in this ecosystem -- particularly focusing on progressive developments in automation and capacity uplift over time. In doing so, we evaluate initial indications of which technological, diffusion, and innovation trajectories the cybercrime ecosystem is following with regard to the adoption and repurposing of GenAI tools.

In practice, we find very little real disruption of the core low-level cybercrime ecosystem by these technologies. Despite a substantial level of interest in generative AI in the cybercrime underground, it has not significantly reduced the skill barrier to entry, nor has it led to serious disruptions to established business models or practices. Instead, its main impact has been on already highly-automated areas such as SEO fraud, social media bots, and some forms of romance scam. We additionally find that most adoption has been of mainstream and legitimate products rather than bespoke tools for cybercrime, and encouraging evidence both that guardrails and economic saturation dynamics are having real effects on preventing the scaling of new forms of harm associated with AI adoption.

\section{Background and literature review -- cybercrime, technology, and generative AI}

\subsection{Cybercrime ecosystems}\label{sec:cybercrimeEcosystem}

In order to understand how GenAI adoption and innovation might take place within the cybercrime ecosystem, we first set out its social and economic dynamics. We focus in this paper on the `cybercrime underground' rather than Serious Organised Crime groups or nation state actors.

 In the 1980s and early 1990s cybercrime was best characterised through individual experimentation and innovation -- small numbers of high harm crimes committed by a tiny population of fairly skilled actors. Forming early enthusiast communities on shared Bulletin Board Forums, individuals would build their own tools, experiment with computer systems, and attempt to break into corporate and government networks, often with surprising degrees of success \cite{jordan_sociology_2017}. This meant that early cybercrime  was largely a matter of small numbers of very disruptive attacks, often involving a great deal of skill, commitment, and experimentation on the part of those committing them. The underground variant of `hacker culture' which characterised this group functioned as a classic subculture with its own values and practice, in particular, valorising technical skill and mastery, anti-authoritarian and libertarian political values, and creativity and experimentation with computers \cite{holt_subcultural_2007, taylor_hackers_2012,turgeman-goldschmidt_meanings_2008}. However, this has developed over the years into a wider and more heterogenous set of distinctive hacker cultures, including fairly mainstream and
non-criminalised variants \cite{coleman_hacker_2008}.

 In the 1990s larger communities of underground ‘hackers’ began to grow, permitted by the rush of users onto the early World Wide Web. These began to resemble an ecosystem of skilled artisans or craft manufacture -- with younger apprentices (‘skids’ in the language of the cybercrime underground) developing skills over time, initially by using tools created by others with little understanding of how they worked, and more well-established elite or `l33t' hackers forming a smaller core of these communities \cite{yar_computer_2005}.

Over time, the economic dynamics of these communities changed. The development of a market for hard-to-use cybercrime tools for sale to end users meant that a substantial secondary market evolved in customer assistance -- helping users to configure and operate these tools. As the tools developed, standardised, and improved economic incentives meant that it became more profitable for producers of tools to package up access to the tool and the supportive labour as a service. Concretely, purchasers would buy services, rather than the tools themselves -- the model known as cybercrime-as-a-service \cite{musotto_more_2022,collier_cybercrime_2021}

As a result cybercrime has undergone something of an industrial transition since the 2000s. With improvements in cybercrime tools, less technical skill was required to operate the different parts of the process. Some forms of cybercrime could be largely automated, with tools being invoked by small numbers of low-waged employees with almost no technical skill. The different steps involved in a given crime could now be split into different industries -- with particular groups specialising in initial access to compromised systems, or renting out access to botnet infrastructure -- and most contemporary forms of cybercrime use pre-made assets and resources circulating online \cite{collier_cybercrime_2021}.

\subsection{The role of technology and innovation in cybercrime economies}\label{sec:technologyUse}

As indicated in previous research, the cybercrime-as-a-service ecosystem has little endogenous capacity for frontier technoscientific innovation~\cite{collier_sophisticated_2022}. New classes of vulnerability or exploits, novel forms of technology, and new high-end red-team practices rarely if ever emerge from the `cybercrime underground', which lacks research and development funding, research and innovation skills, and the equipment or organisation needed to develop new technoscience \cite{collier_sophisticated_2022}. This is mostly the domain of academic researchers, private security companies, and government security services. Even genuinely transformative serious organised crime technology services in both digital and physical contexts (such as the Encrochat phone and successor systems) generally repurpose and redesign existing systems and technologies \cite{guerrero_c_technologies_2019}, and are developed by standalone players seeking to cultivate a niche as an specialist producer serving the wider criminal ecosystem, rather than by established crime groups themselves \cite{bardet_shield_2025,berry_technology_2018}. 

However, the cybercrime underground and organised criminal groups show significant innovation in the \textit{implementation} of technoscientific advances by industry and government, the subversion and adaptation of mature, legitimate technologies and services, and the development of business models, modes of organisation, and effective criminal scripts. On the technical side, most cybercrime makes use of pre-made services, resources, website templates, code, fraud scripts, and technical components that circulate in underground communities. A key barrier to scale for many cybercrime business models is successful automation of the `boring' logistical work involved in administration, cashing out, and customer management, especially in a context in which law enforcement, defenders, and infrastructure providers attempt to make this harder.

\subsection{GenAI adoption in legitimate industry -- tools, capacities, and practices}\label{sec:GenAIAdoption}

In framing how GenAI tools might disrupt the contemporary cybercrime economy, it is instructive to look at their current patterns of adoption in the legitimate economy. As argued in prior work, the use of technology and development of business models within cybercrime generally is much like that of legitimate start-ups~\cite{anderson_silicon_2021}. Cybercrime `start-ups' face many of the same technological, market and organisational pressures as legitimate firms, and cybercrime actors often idolise and borrow from tech entrepreneur culture and practices. However, experimentation in industry has yet to produce conclusive results. A much-vaunted MIT study found that 95\% of business adoptions of GenAI failed~\cite{challapally_genai_2025} -- although this study has been criticised for conflicts of interest, methodological issues, and misleading conclusions~\cite{raynovich_why_2025,gerard_mit_2025}. More recent work presents a complex picture, with some fields capturing early gains, and others finding little success~\cite{fawzy_vibe_2025,becker_measuring_2025}.

Software development has been a notable early use case for GenAI tools which has been a heavy focus in industry. Initially, chatbots such as ChatGPT were adopted by coders as effectively a replacement for cheat sheets and collaborative platforms like Stack Overflow -- answering specific questions about syntax or classic solved problems and producing code that could be copied and pasted into a program. 

It is important to separate the tooling from the model itself, when discussing developments. Larger training datasets, different model architectures, and more parameters can help to improve the model itself, but how the model is used is also important. Initial developments in GenAI for coding could be used to generate scripts, but these would have to be run manually and were typically general purpose depending on the prompt used. IDEs such as Cursor and VS Code with Copilot have improved on this by selecting relevant files and chunks of code, to prepend to the prompt to improve and contextualise generated code. This has been used to help `autocomplete' style features, and generate relevant blocks of code.

Further developments have included agentic coding tools. These are the result of two developments: firstly, models that are fine tuned on generating function calls, e.g. output `WriteToFile(filename, code)' and `ReadFromFile(filename)' in addition to their usual code outputs. Secondly, these models are combined with associated tooling, that prepends requests with relevant files and code chunks, followed by parsing the output looking for such function calls to take action on a codebase. Placing this tooling and model in a loop is agentic coding.

This has given rise to the concept of ``vibe coding'', the use of natural language prompts to produce code, with very little consideration to the user how the code output should be structured \cite{meske_vibe_2025}. Follow-up prompts are then used to fix bugs. This can be useful to get code running quickly, but can result in the build up of technical debt, and the time saved may then used to fix increasingly obscure bugs. This has enabled a lower barrier to entry for non-coders to create their own small programs, from glue code to transform datasets to apps for personal tasks.

This contrasts with the later term of ``vibe engineering'', where the user has experience and knowledge of a suitable direction to take, focusing prompts on small steps to take. Taking smaller steps may help to avoid technical debt build up, and the user takes on a manager-like role with the tooling. Further, as early studies argue, this shifts the nature of programming practice more fundamentally, from a deterministic model to a probabilistic one, and from implementation-based to `collaborative' practices \cite{meske_vibe_2025}.

Legitimate start-ups have, naturally, been major sites of experimentation for Generative AI tool implementation. Existing empirical literature emphasises the importance of \textit{local} configuration of the tools rather than simple generic adoption, and finds that implementation costs for start-ups are often high \cite{donaldson_generative_2025}.

Finally, the major chatbot providers -- and those seeking to implement their technology in downstream products -- have placed a great emphasis on the potential for \textit{agents} to be a major use case for LLMs. This has also lead to increasing development across digital services of APIs for access by LLM agents (and standardised output formats that can be more easily processed by these). Generative AI has also been adopted by businesses for customer support through chatbots on websites, reducing workload for customer support agents on common questions. However, this has risks related to prompt injections if models and tooling do not take steps to limit harm~\cite{shi_lessons_2025}.

Generative AI developments have also taken place outside of text generation: it has been used for image, video, and audio generation. In particular, practice communities \cite{buraga_emergence_2022} have formed not only around experimentation with and sharing of prompts, but more recently around Low-Rank Adaptation (LoRA), a form of retraining that allows fast adaptation and tweaking of models to specific use cases on consumer-grade hardware \cite{hu_lora_2022}.

The distinction in harms between high-effort, specialist adaptation of GenAI tools (which may still cause severe harms at a large volume), versus the harms that emerge where models are made available to the public which enable harmful use cases `out of the box' is an important one. This has been aptly illustrated by the controversy that erupted in early 2026 around the Grok image generator functionality~\cite{murphy_elon_2026}. In this case, the model (explicitly promoted as allowing the generation of sexualised content with very few restrictions, and integrated with the X platform whose owner and `super-user' communities can be argued to deliberately advance misogynistic values) was used to produce non-consensual sexualised and violent images of real people, leading to calls to ban the platform entirely. For our purposes, the level of specialist adaptation and skilled human repurposing required to achieve capacity uplift within cybercrime has important consequences for the nature of the cybercrime phenomena that these technologies might support.

\section{Theory}\label{tech}

\subsection{Innovation theory and evolutionary economics}\label{sec:innovationEconomics}

There is a tendency to make sweeping claims about the ubiquity, inevitability, or desirability of particular models of GenAI use. However, technological innovation by providers, adoption by users, and disruption of ecosystems are not simple, inevitable, or homogeneous processes. Choices are involved -- technology created by producers is generally taken up piecemeal and tentatively by users, who experiment with early forms of use, adapt the technology to their local environment, and develop their own innovations. Early effective implementations in one domain may simply be copied in a range of other domains -- thus, we may observe path dependence for early successes and incomplete adoption rather than a full exploration and optimisation of the possibility space). Similarly, as repeatedly seen in studies of the software industry, new innovations frequently take a long time to be tailored usefully to local contexts \cite{pollock_software_2008}. What we see instead is an often slow, piecemeal, and back-and-forth journey of coupling of practices, technology implementations, and knowledge between producer and user markets.

Thus, contemporary perspectives from innovation studies understand innovation as happening in an array of local niches of development, experimentation, and use case implementation. As these niches are subject to a selection environment, promising or successful paradigms emerge and mature -- these give rise to trajectories of innovation as early features emerge and are implemented, with numerous failures \cite{dosi_technical_2010}. 

The idea of the technological trajectory has also been taken up within the cybersecurity literature  \cite{anderson_silicon_2021,kraemer-mbula_cybercrime_2013}. Anderson et al argue that the use of technology in cybercrime ecosystems can be understood as a process of entrepreneurship -- functioning in much the same way as technology adoption in the legitimate start-up ecosystem. Technological innovation in cybercrime ecosystems follows trajectories that bind together experimentation and innovation with the development of business models, pathways to scale, and wider organisational effects~\cite{anderson_silicon_2021}. Anderson et al. theorise these trajectories through a linear curve model with six stages, in which a form of criminalised technology use initially requires (1) a set of \textit{preconditions} that enable the trajectory to begin; (2) a number of barriers to entry (such as skill, or the availability of particular resources) that must be overcome; (3) a set of potential pathways to scale beyond initial experimentation that entrepreneurs compete to explore; (4) a number of bottlenecks that can inhibit scaling; (5) a relative absence or frustration of defenders; and finally (6) saturation mechanisms (including competition over resources) that establish hard limits to growth.

Compared to the legitimate technology ecosystem, the selection environment for criminalised technical innovation is far harsher, with no institutions protecting nascent innovations from market forces or predatory competitors, weak mechanisms of information diffusion, and no major sources of external capital investment or leverage. In addition to increased risks of business failure, law enforcement (and competition from an overheated legitimate tech sector) provide further selection effects, often effectively draining talent pools and restricting demand. There is little economic incentive to pursue new innovations and as a result, genuine innovative disruptions are extremely rare. As a result of these factors, the trajectories we do observe take a long time to mature, and many experiments fail before scaling. 

In addition to these economic dynamics, the cybercrime underground also has a distinctive cultural relationship to technology that shapes the dynamics of use and adoption. As argued by Silverstone in work on the \textit{domestication} of technologies by users, innovation and implementation not only involve the incorporation of technology into economic activity, but into a moral economy \cite{silverstone_listening_1991}. Users make sense of new innovations by fitting them into their existing cultural universe of meanings, values, and practices. \cite{sorensen_domestication_2006,silverstone_listening_1991, haddon_roger_2007, hirsch_information_2003} and these directly shape how they adopt. In exploring adoption, we therefore draw both on the innovation dynamics of evolutionary economics, but also consider the hacker subculture's understanding of technology and skill in their own communities. Given the centrality of experimentation and `opening the black box' in traditional hacker culture, and the importance of technical skill in structuring social capital within the hacker field \cite{coleman_hacker_2008}, we would expect GenAI technologies in particular to be both an enticing new development, but also potentially a disruptive threat to established social order and shared identities. Therefore, our findings discuss dynamics of both saturation and scale; both defensive resistance and wild enthusiasm; at play in the cybercrime subculture. 

\section{Prior work and likely innovation trajectories}

Before turning to our empirical research, we set out a series of hypothetical adoption trajectories for GenAI tools in the cybercrime underground, and coin some novel terms with which to describe potential use forms that might arise from differing modes and levels of adoption. We note that these are not simply technological forms -- they combine business case, revenue model, organising structure, working practices, and wider relationships to the overall cybercrime ecosystem with particular forms of technological adoption and innovation. 

In this section, we also set out and organise the current research literature on AI use in the cybercrime ecosystem, organised around these trajectories. The current literature on the use of GenAI tools in cybercrime is largely speculative. Many papers present solely theoretical models of potential disruption that might be expected \cite{choi_understanding_2024,syed_ai-powered_2022,garg_artificial_2024,abid_cybercrime_2024,karamchand_detecting_2025,zucca_regulating_2025,treleaven_future_2023}.  A further set of papers analyse case studies taken from the cybersecurity press or reports from security providers \cite{phillips_darwinian_2025, aro_use_2025, liu_network_2024,ionescu_generative_2025}. As is argued in the cybersecurity literature \cite{collier_not_2025}, while useful in illustrating potential effects, these are far from objective sources of primary data, as there are clear incentives to present small-scale experimental adoption as a major phenomenon. We can find three studies that uses empirical data from forums. The first \cite{shetty_investigating_2024} collects some discussions of prompts found circulating in the cybercrime underground, though most of the analysis is based on law enforcement interviews. The second, \cite{dupont_what_2026} uses private sector data scraped from forums by security companies to study some early elements of diffusion. Analysing 160 threads, they find that the forums they study largely focus on attempts to jailbreak or adopt mainstream AI tools, with very little successful adoption of custom-made `Dark AI' products (this aligns well with our own findings). Much as we do in this paper, they similarly find a purely linear model of diffusion does not adequately explain the apparent scepticism of much of the cybercrime underground. A third study by Lin et al.~\cite{lin_malla_2024} look at LLM jailbreaking as a service, exploring 212 services based upon eight backend services. The authors gain access to these services, and use these to generate malicious code, phishing emails, and phishing websites. They note that malicious code generation can generate basic exploit payloads, and test phishing emails and websites against users on Mechanical Turk. We note the work by Lin et al. was carried out in 2023, and models, tooling, and guardrails will have progressed considerably since this prior work. 

Other studies focus entirely on AI adoption in defence \cite{alansary_emerging_2025,singh_is_2025}. A further body of useful work uses interviews with law enforcement and other experts \cite{qiu_policing_2025,channapattan_ai_2025, burton_ai_2025} or surveys of the general public \cite{alawida_unveiling_2024}. One example compares some quantitive data -- abusive IPs and and crypto scam reports -- before and after the release of ChatGPT \cite{luu_unintentional_2025}, reporting an increase in these measures. There are a small number of important studies to date which develop insights into the early signs of adoption (for example, \cite{burton_ai_2025}), though these necessarily focus empirically on what is being observed by defenders, rather than on direct empirical study of communities of attackers.  We can find no \textit{systematic} or community-scale empirical studies of Generative AI adoption in the cybercrime underground to date.

The largest vein of academic research involving data uses simulation methods, in which security researchers jailbreak models and experiment with their capabilities in lab settings \cite{usman_is_2024, alotaibi_cyberattacks_2024, teichmann_ransomware_2023,hu_exploring_2025,gupta_chatgpt_2023}. We argue that as interesting and potentially useful as lab-based jailbreaking simulations or theoretical research are, they are a poor guide to actual adoption pathways and utility in the cybercrime underground (for much the same reasons that highly complex attacks like Rowhammer are largely not a feature of cybercrime in the wild). What is achievable experimentally in a well-resourced academic or corporate computer lab by experienced researchers is not the same as what is practically available to those in the cybercrime underground -- and theoretical implementations even where demonstrated as feasible in simulations may not in fact greatly improve business models or cybercrime practices in practice. Economic and social dynamics mediate adoption and implementation -- systematic empirical work is needed to establish what early use cases are \textit{actually} being exploited at scale (beyond eye-catching individual case studies that may simply show experimentation with a new tool that is never turned into harm at scale in practice). Additionally, costs for use of these new technologies may in fact exceed those of simply using older and simpler but more `mature' technologies that are well integrated into business models -- such as simply using pre-made scripts.

Finally, the AI platform companies \cite{anthropic_full_2025,noauthor_adversarial_2025} and security vendors have themselves published a number of reports using their own data, in which the data is generally not made publicly available.

We therefore now set out, based on the current conditions of the cybercrime ecosystem and the existing literature, a number of areas of GenAI innovation which appear to be particularly promising areas of potential adoption. These are intended to structure the discussion of our empirical findings -- to allow us to compare the tentative initial adoption and innovation trajectories we observe with these hypothetical cases, and identify any current `reverse salients' that appear to be blocking particular trajectories at present. 

\subsection{Vibe hacking}

A major hypothecated adoption trajectory involves so-called `vibe hacking', namely, the use of vibe coding and vibe engineering approaches by semi- and high-skilled actors to improve or speed up development tasks, system penetration work, and other cybercrime practices. This has largely to date been observed through individual eye-catching case studies, almost always in nation-state or nation-state-linked contexts. An Anthropic paper claims observed use of coding assistants by groups involved in cyber-espionage, but not for innovation - rather, for `automation of orchestration of commodity open source tools' in order to achieve greater scale \cite{anthropic_full_2025}. A rebuttal article however suggests a high failure rate for these use cases and reflects general scepticism that reported efficiency gains are making a significant difference  \cite{goodin_researchers_2025,burton_ai_2025}.

\subsection{Creation of new markets and targets}\label{sec:marketsTargets}

 Many of the primary impacts of AI on cybercrime are expected to initially come from the mass adoption of these systems by legitimate industry -- and hence changes in the characteristics of targets -- rather than primary innovation within cybercrime ecosystems. Early adoption of chatbots by businesses gave rise to a number of reports of users being able to convince them to provide data or free services through `prompt injection' attacks (not conceptually dissimilar to SQL injection vulnerabilities) \cite{shi_lessons_2025}. Other reports suggest that as companies adopt internal LLMs for their own business information systems, these will provide new mechanisms for privilege escalation and access to secure data once attackers gain access \cite{oesch_living_2025}. 

More generally, the use of LLMs in coding may create major new attack surfaces as poorly-written and vulnerability-rich code proliferates in services. Similarly, `hallucination squatting' attacks suggest further potential vulnerabilities might be created by the legitimate adoption of coding assistants. Finally, there is early speculation that adversarial attacks on chatbot platforms themselves -- either through prompt injection or through manipulation of training data -- may be an area of deviant innovation. This itself has given rise to a number of start-up security firms offering expertise in mitigating such manipulation attacks.

Where we do expect to see genuinely novel innovations, these are likely to concentrate in areas with high R+D resource and clear incentives - for example, nation state security services and some academic labs or cybersecurity companies. As in the cybersecurity ecosystem, however, these face strong barriers to diffusion into active use `in the wild'.

\subsection{Logistics, automation, and organisation}\label{sec:automation}

Although changes to coding or development practices in cybercrime have been a focus of commentary, these are in fact fairly marginal skills in the cybercrime ecosystem. Instead, logistical, administrative, organisational, and entrepreneurial skills play a larger role in success for cybercrime-as-a-service businesses, and hence, in volume cybercrime~\cite{anderson_silicon_2021}. These logistical improvements and tools are also where much of the active resources are being focused for AI innovation and implementation in the legitimate economy. Many of these tools and products will either have direct dual use capability in cybercrime, or simply have uses in solving the kinds of organisational problems that all businesses -- including illegal ones -- face. Thus we may see better project management, finance, HR, and server administration tools being adopted to make cybercrime-as-a-service business models more efficient and more automatable at scale (as reported in \cite{burton_ai_2025}). Similarly, changes to infrastructure, tools, and practices in the advertising economy will further affect criminal businesses, many of whom use social media influencers and paid advertising to attract customers. Better translation services could additionally support an extension of labour, customer, and victim markets, potentially leading to an internationalisation of cybercrime markets (which are currently heavily structured by shared language).

\subsection{Social engineering}\label{sec:socialEngineering}

Social engineering, fraud, scams, and harassment are all areas where defenders, the public, and law enforcement report GenAI adoption by attackers \cite{burton_ai_2025}. Experimental use cases have been observed in automating aspects of social engineering, particularly in high-visibility case studies. There are notable improvements being observed in spam and scam scripts, particularly in the quality of the writing and translation. Clear use cases in video and audio synthesis can be observed\cite{burton_ai_2025}, though again the results at scale of these experimentations on business models are unclear.

\subsection{Jailbreaking and other restriction-bypassing}

As essentially no actors in the underground ecosystem have the capacity, resources, or technical knowledge necessary to carry out frontier LLM development, these trajectories are likely to be built on a technical base of adopting and adapting mainstream and legitimate technologies developed by major tech players or open source research communities. Given that (with the possible exception of platforms built for the cybersecurity industry) these are not optimised for cybercrime use cases and are often actively guardrailed against use for this purpose, a focus of innovation for this trajectory is likely to be on \textit{jailbreaking} models, on \textit{retraining} and repurposing them (for example, using the kinds of LoRA practices we see in enthusiast communities), or developing OPSEC practices to allow LLM use for innocuous parts of criminal enterprises without triggering suspicion and a flag to law enforcement. Scholarship to date notes the importance of open-weight models as less effectively guardrailed and more prone to abuse \cite{burton_ai_2025}

We already observe this in other areas of crime. An ARU study on CSAM and LLMs~\cite{davy_artificial_2024} finds that abuse communities are retraining the models using LoRAs, both for adaptability to use context and to get around restrictions at source. More widely, there are growing legitimate, quasi-legitimate, and criminal communities developing and sharing their own practices at repurposing LLMs, often in the form of prompt-sharing and trading retrained models. These communities exhibit the social richness, burgeoning body of practices and developing habitus to constitute an increasingly stable set of subcultures of their own (though often derided and rejected by the established practice subcultures in, for example, non-GenAI pixel art, electronic music, and the like) \cite{buraga_emergence_2022}. We would expect to see this kind of social tension in the cybercrime communities as well -- as established subcultures facing disruption by new technologies and their enthusiasts. However, the centrality of technology, experimentation, and subversion of technological design to the underground `hacker' subculture suggests that this may play out differently in this case.

\subsection{AI and defenders}

In the cybersecurity industry, a great deal of the work involved in actively responding to threats (rather than securing systems preventatively) is based on pattern recognition. The automation and re-use of materials and resources involved in volume cybercrime means that characteristic `signals' can be detected, searched for, and trigger mobilisation of a defence (either automated or manual). Current tools effectively search for current well-known patterns and characteristic anomalies - as resources like mass phishing scripts, fraud website content, and spam are generated through regular expressions, they can be tackled through pattern recognition. However, GenAI tools allow far more variability and `human-like' content, resources, and activity patterns to be generated, making these approaches less effective.

\subsection{Social and cultural effects on the cybercrime underground}

Further, these potential forms of adoption have potential consequences for the \textit{culture} of different parts of the cybercrime underground. As observed with previous industrial transitions in cybercrime, the material political economy of work is closely related to the experiences and motivations of the people who do it, and hence, to how they understand and value what they do \cite{collier_cybercrime_2020, coleman_hacker_2008}. This cultural element plays an important role in understanding cybercrime. For example, earlier work has observed that the move to cybercrime-as-a-service was one of a number of factors contributing to a deskilling of work in the cybercrime economy, which displaced the traditional `hacker' ethos of individual experimentation and mastery in favour of a more entrepreneurial model which ultimately left many of the workers underpaid, bored, and overworked \cite{collier_cybercrime_2021}.

\subsection{Maximal and minimal cases for industrial transformation}\label{sec:transformation}

In our introduction, we set out two potential `futures' of GenAI disruption of cybercrime as an industrial ecosystem, constituting the maximal and minimal cases. Before turning to our methods and findings -- the early signs of what is actually happening -- we outline these hypothetical scenarios here in detail.

\subsubsection{Stand-alone complex} This can be considered the maximal case that might be expected from the current state of the cybercrime ecosystem and the potential capabilities of LLM tools currently in development and use -- namely, the development of autonomous or semi-autonomous crime-gang-in-a-box infrastructure. In this scenario, the adoption of mature AI agent technologies leads to a full-scale industrial transformation of the cybercrime ecosystem of the kind seen in the rise of the cybercrime-as-a-service model. This would co-occur with changes to the work, culture, practices, tools, infrastructures and economic dynamics of cybercrime. 

For ease of discussion, we term this the \textbf{`Stand-Alone Complex'} scenario \footnote{With reference to the popular anime series Ghost in the Shell, in which the increased digitisation of society and autonomy of networked computing systems is depicted as leading to inexplicable and terrifying mass crime and terror campaigns emerging in digital society seemingly from nowhere -- a `copy without an original'~\cite{baudrillard_simulacra_1994}.} We use this term to encapsulate a range of (entirely hypothetical) scenarios in which semi-autonomous agentic AI technologies become a major set of actors in the cyber-risk landscape. This scenario is not intended as a prediction, rather it is a sensitising lens through which to explore existing early trends in at-scale automation in cybercrime. It usefully foregrounds the economic dynamics of such scenarios -- for example, where they may trend towards saturation or resource exhaustion, and where genuine harm at scale may become economically practicable given often fairly small innovations in digital automation.

It is important to note (as we describe above) that most forms of cybercrime are already either highly automated, use largely trivial pre-made resources (such as code, scripts, website templates, and images), or both. This scenario -- a step-change in automation -- would likely not \textit{entirely} remove the need for human labour in the commission of cybercrime (not least due to the unreliability of jury-rigged tools, the cat-and-mouse games that might be expected between the underground, law enforcement and tech companies, and the competitive dynamics between different illicit actors), but would further stratify and deskill administrative and maintenance  work in the ecosystem. For many forms of cybercrime, particularly Denial of Service, botnet management, and Ransomware, there would be immediate strategic competition for resources (e.g. infected machines, reflectors, spoofing sites, servers, hosting and payment infrastructure) and customers between players. Especially in the case where automation permits quicker and higher scale, these competitive dynamics could be expected to escalate conflict (and technological innovation) within the cybercrime ecosystem. The wider cultures of cybercrime might also be expected to change -- for example, heralding a move from the increasingly deskilled `cybercrime as a service' model to an ecosystem that once again privileges technical skill, hard mastery, and individual acclaim, but with the valuable individual `hacker' developing AI wrangling skills rather than learning to craft payloads or engineer buffer overflows.

\subsubsection{Vibercrime} At the other extreme, we posit a possible scenario in which GenAI technologies are adopted piecemeal by both low- and high-skilled actors, largely to achieve basic automation and coding support, but the organisation of the ecosystem, its technological base, and the core business models of cybercrime remain largely the same. In this scenario, highly-skilled actors will see modest productivity gains as they adopt industry standard IDE tools, balanced out by the need to check and audit the code produced in more depth. This may lead to the achievement of larger scale for some criminal organisations, or simply to minor labour changes in the market as single coders can maintain and develop a far larger number of projects simultaneously, so can work across multiple organisations. For the lower-skilled actors, this may lower the skill barrier to entry enough to bring more people into the ecosystem -- creating a glut of low-quality offerings running on fragile codebases that their maintainers do not understand, and which are full of security vulnerabilities. 

Paradoxically, this might result in similar cultural consequences to the Stand-Alone Complex case. Namely, this could be expected to also bring the solo hacker back to cultural prominence, creating a new ecosystem of script-kiddies attempting to wrangle their code bases into working order and a highly valuable class of effective AI-supported coders.

Drawing on the language of `vibe coding' and the misleading term `vibe hacking'\footnote{This is a misleading term as most of the work being done by the LLM here is not `hacking' in any sense, rather it is fairly basic programming or software engineering \cite{collier_sophisticated_2022}}, we term this the \textbf{`vibercrime'} model -- one in which the proliferation of coding assistant tools and automation of basic admin tasks which were beyond the skills of the cybercrime underground leads to first a drop in the skill barrier to entry, and then a drastic reduction in the number of staff needed to successfully start up and scale a cybercrime enterprise. This also implies the rise of the so-called \textbf{vibercriminal} -- whose skills lie not in crafting payloads or reverse-engineering machine code to achieve buffer overflows, but wrangling and coaxing AI systems for attack, defence, and criminal logistics.\footnote{Names are important in cybercrime~\cite{hutchings_amplification_2024}. We have here deliberately chosen a rather pompous name taken from science fiction to signify the importance of Science Fiction imaginaries of AI in shaping these wider risk perceptions, and a second, faintly ridiculous name to emphasise the low-rent nature of much cybercrime in practice. Our intent here is both to exemplify the characteristics of each of these cases, but also to demonstrate the importance of terminology and marketing in shaping how people interpret claims made by researchers about cybercrime.}

\section{Methods and research questions}\label{sec:methods}

In exploring the early adoption of GenAI tools by the cybercrime ecosystem, we draw on a range of data sources and methods, which we discuss in this section. We largely adapt methods from previous research on underground cybercrime forums~\cite{pastrana_crimebb_2018,vu_turning_2020,pete_social_2020,hughes_art_2024,hughes_playing_2019}, though we additionally experiment with some novel techniques, detailed below.

\subsection{Research Questions}

Our study is exploratory, but driven by the following research questions:

\begin{enumerate}
    \item \textit{How much interest is the underground cybercrime ecosystem showing in these tools and what effects are they having on  cybercrime subcultures?} What are the general impressions of these tools in the cybercrime underground, and how favourably do these communities view their use?
    \item \textit{Are these communities adopting these tools, and if so, how? }Can we observe any initial successes and failures in local innovation with these tools for cybercrime-specific use cases? What innovation trajectories can we observe?
    \item \textit{What are the economic effects of GenAI innovations on cybercrime? }Are they reducing skill barriers to entry for neophytes, or producing skill/productivity multipliers for experienced users? Are they assisting new or existing users in learning cybercrime skills? Do we observe saturation?
\end{enumerate}

\subsection{Data Sources}

We use data from CrimeBB, from the Cambridge Cybercrime Centre. \cite{pastrana_crimebb_2018} This dataset contains posts scraped from underground and dark web forums on the topic of cybercrime. Underground forums are public forums that are not found on the dark web, and dark web forums are only accessible through anonymity networks such as Tor. The collection includes a range of different forums, from general cybercrime discussion forums to those dedicated to more specific topics, such as account compromise, SEO fraud, videogame cheating, passive income generation and romance scams \cite{pastrana_measuring_2019}.

The dataset is structured hierarchically, with sites which contain boards, boards which contain threads, and threads with contain posts. Threads are a series of on-topic posts, ordered by time, and the topic is set by the first poster. Boards are more general categories, collating groups of threads.

We select a sample from this dataset, aiming to collect a large sample to reduce down later. We create a word list to search by. This includes (case-insensitive): ``artificial intelligence'', ``LLM'', ``GPT'', ``Claude'', ``Gemini'', ``prompt'', ``Copilot'', ``vibe coding'', ``OpenAI'', ``model'', ``generative'', ``machine learning'', ``AI'' (uppercase only). 

We use these word lists to identify: boards with matching board names, threads with matching thread names, posts with matching keywords. Where a board name matches, we include all threads and posts within that board. Where a thread name matches, we include all posts within that thread. Where a post's contents match, we include the entire thread containing this post. This does include threads with just one relevant post, which we decide to include to gather all relevant threads to later filter and discard if necessary.

We then create ``documents'' (entire threads) for analysis: each thread becomes a single text field consisting of the thread title, followed by each post contents in order. This is approach is chosen to collate related posts to retain contextual information, as thread replies may miss context without the title and first post.

This sampling method obtains 808,526 threads, up to a cut-off date of 2025-12-10 when the sample was taken.

We reduce this initial sample by removing threads before the release of ChatGPT in November 2022. Only threads started on or after 2022-11-01 are retained. We assume forum moderation norms in this sample, namely where new topics are required to have their own threads, rather than contributing vibe coding-related discussions to a long running historic thread. 

The final sample contains \textbf{97,895 threads} for computational and qualitative analysis, between \textbf{2022-11-01 and 2025-12-10}. We note the large reduction is due to the removal of threads prior to the release of ChatGPT.

\subsection{Computational methods}
We initially take a computational quantitative approach, to explore the large sample gathered, as it would be infeasible to read all threads. Furthermore, this allows us to explore the dataset through topic categories and changes over time to later contextualise the qualitative results. 

We fitted a hierarchical topic model, using BERTopic with HDBSCAN~\cite{grootendorst_bertopic_2022}. We initially start with a minimum of 10 threads per topic, in which 10 was selected as it returned the most coherent topics qualitatively, using the embedding model all-MiniLM-L6-v2. The number of threads per topic is selected to balance between a small number of topics, which may be too general, with a large number of topics adding additional challenge to analysis. 

The model produced 122 topics, and a ``-1'' outlier topic in which threads were not able to be grouped by HDBSCAN. Our approach using BERTopic with HDBSCAN will try to place threads into set topics, and where there are not a minimum of 10 related threads for a topic, these are placed into an outlier topic. While other approaches such as Latent Dirichlet Allocation (LDA) can provide ``topics'' to all threads, these are assigned through a document having of a distribution of topics. We choose to use BERTopic, as this uses embeddings to provide a suitable representation of threads instead of the word occurrence and co-occurrence used by LDA. It is possible to put these outliers into topics, such as by identifying nearest topic neighbours, however this can make topics less coherent. We note the limitations of using this approach, including that discussions on evolving and new AI topics may lead to small novel topics, leading to outlier topics.

Validation of topic outputs is important to understand if a topic model is a good representative fit. We identified a set of six AI-themed boards for one site in the dataset, as this provides ground truth to compare the model to. We found 43.7\% of threads within these boards were classified as an outlier. While this has limitations, we are able to compare changes over time for identified topics, and we complement computational quantitative findings with our qualitative approach.

\subsection{Qualitative methods}

In addition to the computational and quantitative analysis, we conducted extensive qualitative analysis of these collections. We did this by combining three different sampling approaches to create a number of qualitative samples for analysis. 

Approach 1 used a full random sample of the data, stratified over time and forums, producing 3 (non-overlapping) qualitative samples of 100 threads each, for a total of 300 threads. 

Approach 2 used topic model stratification as a first stage to the random sample (using the topic model described previously). This produced 3 samples with 100 threads per sample, for a total of 300 further threads. 

Approach 3 used a local LLM model, \textit{openai/gpt-oss-20b}, to classify the threads in relation to vibe coding and vibe hacking, as well as identifying threads where users had discussed changing practice and overcoming barriers. We later find the categories classified did not individually produce relevant results (specifically the overcoming barriers and changing practices labels), so we combined all non-other labels to examine all vibe-coding-related labels as one for this approach. We include our original method and prompt to document our approach. We select our prompt for classification by adding a small amount of context, along with a short classification guide and instructions to select only one or ``other''. We include ``somewhat related'' and ``other'' in the prompt to recognise not all threads will be related, and instruct the LLM that ``other'' is an acceptable label. We run this on an initial sample of 100 threads, and manually check the results, which we found to be acceptable. Note there were less than five overcoming barrier and changing practice labels in this initial sample, and the limitations of these two labels were not apparent until classifying all over a time period of four days \footnote{We invite others to critique our experimental use of LLMs here, both from an ethical standpoint (i.e. the position that these technologies are too compromised for any forms of use), and from a practical standpoint (i.e. `you should learn to prompt better'). To the first, we would argue that recognising the manifest ethical issues with much LLM use, we have adopted a series of mitigations here detailed above, and feel that to write about generative AI adoption in a technological ecosystem in an open and thoughtful way, it is useful for us to develop further familiarity with the technologies themselves, their affordances, capacities and cultures. To the practical critique, we contend that it is non-trivial to repeatedly and randomly alter the research instrument with no directed sense of how or why this is affecting the non-deterministic mechanistic elements of the generated sample and classification. While iterative instrument refinement is common in both qualitative and quantitative approaches, this invovles a level of \textit{reflexivity} and \textit{mechanistic assesment} that enables the truth claims made to be defended - something that is not possible with a `prompt engineering' approach. In this case, we have used our LLM sampler (fairly arbitrarily) in combination with more directed methods to generate a sample for rich qualitative reading.}. The prompt used was: 

\begin{lstlisting}
You are an expert qualitative researcher classifying cybercrime forum discussion documents into predefined themes.

These are all somewhat related to AI, vibe coding, vibe hacking, and cybercrime. We want to find those discussions related to changes, including experiences of using them.

We want to look at how LLMs and vibe coding and vibe hacking have changed cybercrime.

Your task is to identify which predefined theme best matches each document. Use priority theme where you see some discussion of this. 

Available Themes (YOU MUST SELECT ONE FROM THIS LIST):
- overcome_barrier: Specific discussion that vibe coding/hacking has personally helped them overcome an issue or barrier (priority theme)
- change_practice: Specific discussion that vibe coding/hacking has changed how they operate (priority theme)
- using_vibe_coding: Discussion about personally using vibe coding/hacking or related, that isn't suitable for change_practice or overcome_barrier
- about_vibe_coding: Discussion about vibe coding/hacking as a phenomenon, rather than using personally
- other: Documents that don't clearly fit into other categories

Guidelines:
1. Select the SINGLE theme that best matches the document's primary focus
2. You MUST choose from the themes listed above - no other values are allowed
3. If the document doesn't clearly fit any specific theme, select ``other''
4. Base your decision on the document's main content, not peripheral mentions
5. Consider the semantic meaning and context, not just keyword matching
    
\end{lstlisting}

Of the total 97,895 threads in the sample for analysis, 95,292 (97.3\%) were categorised as other. 1,821 threads (1.9\%) were categorised as using vibe coding, and 432 threads (0.4\%) were categorised as about vibe coding. 244 threads (0.2\%) were categorised as changing practice, and 106 threads (0.1\%) were categorised as overcoming barriers.

For qualitative analysis with Approach 3, the 2,603 threads of non-other categories were grouped together and used as a single sample.

The overall qualitative sample, combining all three approaches, contained \textbf{3,203 threads} out of a total of 97,895 AI-relevant, post-ChatGPT launch threads. These were analysed qualitatively by two coders. To further develop the immersive ethnographic component of the analysis, one of the coders read and analysed an additional more than 20,000 discussions selected evenly from across all three samples, and spent around 5 hours accessing and reading the scraped sites directly in situ.

The analytical approach for this qualitative work was general thematic coding, involving reading and coding the discussions inductively. Themes were developed initially from an inductive approach, then clustered into higher level themes as the coding progressed. These high-level themes are largely replicated in the structure of the qualitative section of the findings chapter.

As this coding involved detailed inspection of the posts themselves at scale, we are able to discuss the accuracy of the LLM classifier. We report only rough impressions here as this aspect of the study design was simply to experiment with different sampling techniques for qualitative analysis rather than to produce robust quantitative measures (which we do produce for our computational study). Through qualitative analysis, we find roughly 80\% of posts classified as positive (i.e. not ``other'') by the LLM were relevant to discussing AI and vibe coding, with around 2 in 10 threads completely unrelated. The classification of individual labels was almost universally inaccurate -- the LLM failed on almost every post to identify the category correctly (overcome\_barrier, change\_practice, about\_vibe\_coding, using\_vibe\_coding) and so we do not analyse these categorical results quantitatively. We acknowledge the LLM approach can be improved with fine-tuning on ground-truth data, and running experiments with further changes to the prompt, which we leave to future work.

The threads included in the LLM-based sample were on average much longer, more detailed, and more obviously relevant to LLM use and vibe coding on first inspection. However, if taken as a single dataset instead of using multi-approach sampling, they give a wholly misleading characterisation of activity compared to the random and topic samples. They overwhelmingly represent the in fact fairly small corpus of discussions of vibe coding (as that is how we structured the prompt), while the traditional topic model and random samples give a far more heterogeneous view, surfacing forms of LLM use and practices that do not appear in our LLM-selected dataset at all, but which are on analysis far more prevalent. As we discuss in our findings, and as the cybercrime actors themselves have found, we find that in this project LLM tooling can largely only give you what you already know to ask for, and this can blunt their utility as an exploratory methodological tool for social data science.

\subsection{Data post-January 2026}
While this paper initially focused on data up to and including 2025, we later take an additional sample of data, up to 2026-07-10, to explore changes that occurred during the first half of 2026. Between 2022-11-01 and 2026-07-10, this sample for quantitative and qualitative analysis included 175,290 threads.

\begin{figure*}
    \centering
    \includegraphics[width=0.95\linewidth]{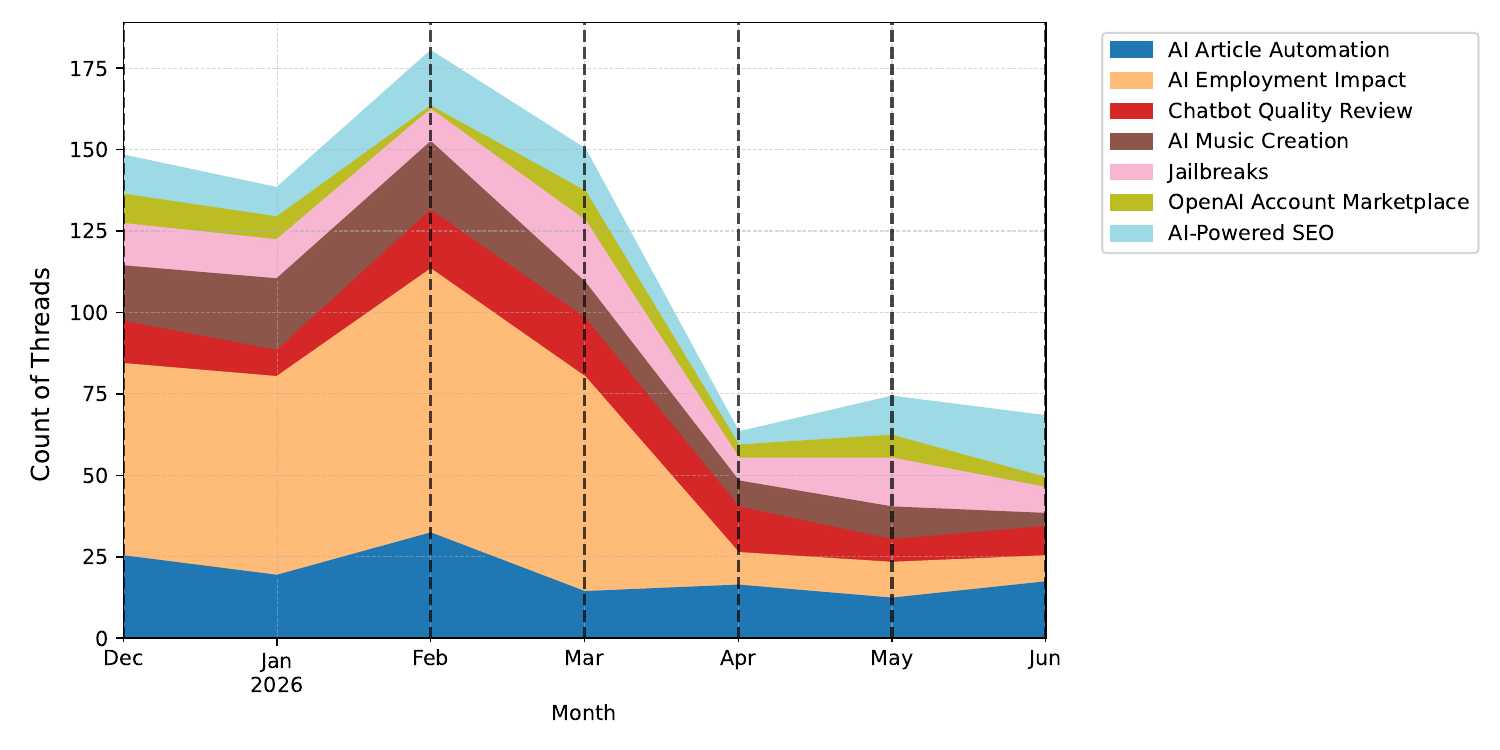}
    \caption{Stacked AI-Related Topics Over Time After December 2025 (Excluding Outliers)}
    \label{fig:topics_over_time_stacked_2026}
\end{figure*}

\begin{figure*}
    \centering
    \includegraphics[width=0.8\linewidth]{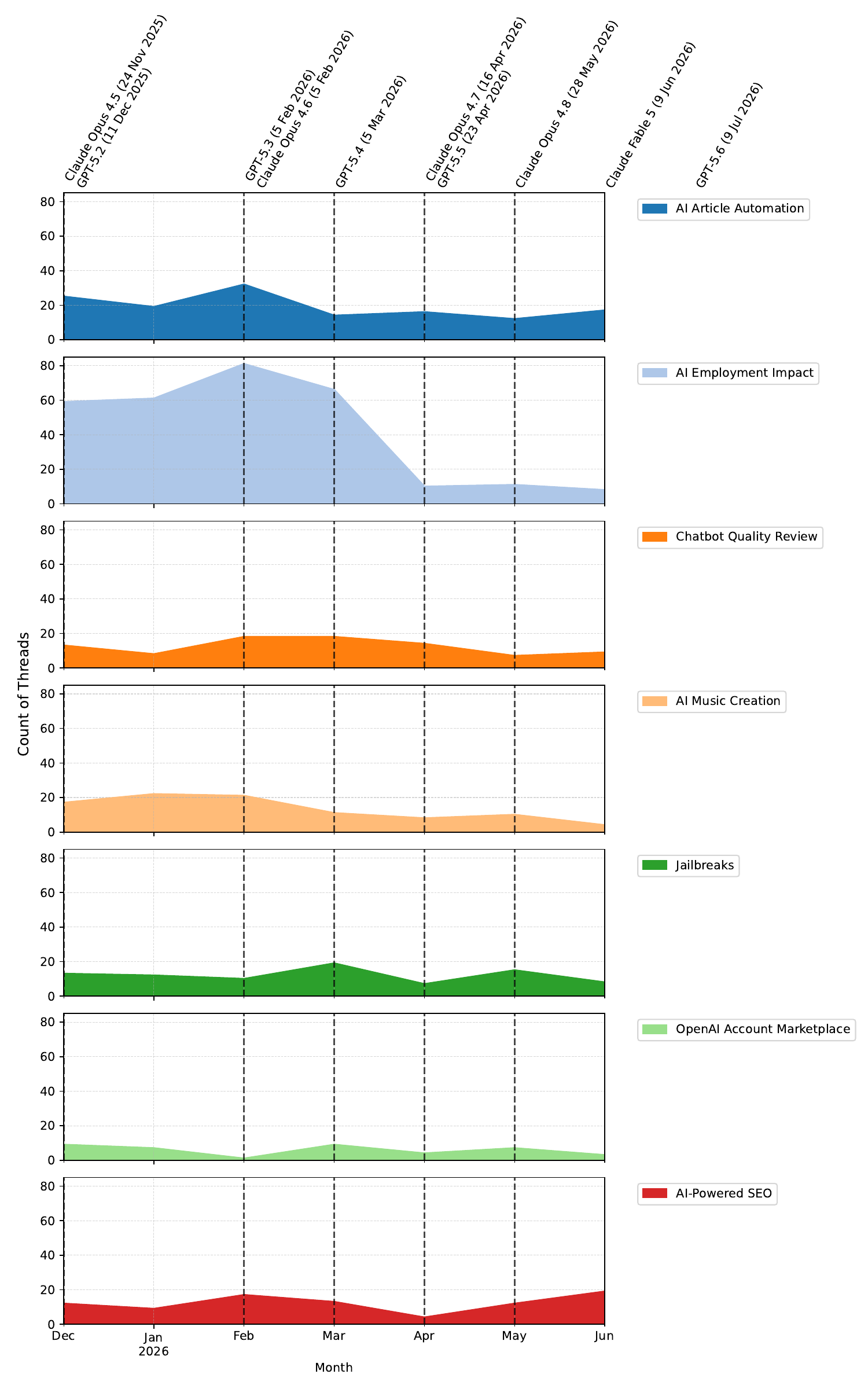}
    \caption{AI-Related Topics Over Time After December 2025 (Excluding Outliers)}
    \label{fig:topics_over_time_2026}
\end{figure*}

We fit a new topic model to this sample, which produced 142 topics as well as an outlier topic. Figures \ref{fig:topics_over_time_stacked_2026} and \ref{fig:topics_over_time_2026} show the AI-related topics over time for 2026.

We re-run LLM labelling of the sample, and produce an overall sample of 575 threads for qualitative analysis. This includes 548 threads from LLM labelling, 100 from keyword sampling, and 100 from stratified sampling, with the final sample using the union of these three.

We discuss our qualitative results in section \ref{sec:developments_post_2026}.

\subsection{Ethical considerations}

For this project, we observe a collection of cybercrime forums, analysing discussions around AI quantitatively and qualitatively. This requires careful consideration of the stakeholders, possible harms to them, and precautions taken to minimise harms. While CrimeBB contains data on cybercrime discussion, the project is not looking to tell if crimes have been committed using generative AI, and is representative of only a particular demographic of cybercrime communities. This project explores perceptions and discussions of usage of generative AI, taking place on cybercrime forum platforms. The project was granted ethics approval by the department's Ethics Committee. 

\textbf{Stakeholders and harm considerations}. We identify a number of stakeholders, possible harms to them and mitigations to minimise this potential harm.

\emph{Participants in the forums}. Individuals actively involved in the forums may risk legal repercussions if identified. Even perceived association with illicit activity could lead to consequences, and the expectation of anonymity and pseudoanonymity within these forums is a significant concern. We do not attempt to de-anonymise any forum participants at any point, and do not name forums. All quotes are altered from their original form. We stress that not all discussions in these forums are related to illicit activity and crime, as community building and general chatter are all part of these communities.  

\emph{Cybercrime forum platforms}.
Public exposure of the prevalence of these forums could attract increased scrutiny from regulators and law enforcement. This may lead to demands for greater content moderation and takedowns, potentially impacting forum users more broadly.
Furthermore, we emphasise that while discussions analysed in this study are from cybercrime-focused forums, their connection to illicit activity is uncertain, and outputs do not and should not be interpreted in this way.

\emph{Researchers \& Institutions}. Exposure to disturbing content carries the risk of psychological distress and secondary trauma for the research team. Further, there is a potential, albeit low, risk of retaliation of harassment from individuals involved to both researchers and institutions.
Researchers involved in the research had access to mental health resources or debriefing sessions to address potential psychological impacts from exposure to disturbing content.

\textbf{Other considerations}. The CrimeBB dataset was provided through data sharing agreements. 
We could not gain informed consent from all members of the observed forums. According to the British Society of Criminology's ethical guidelines\cite{criminology_statement_2015}, informed consent may be waived for research examining publicly accessible online communities, particularly when the focus is on collective patterns of behaviour rather than individual actions. 

We do not foresee any negative impacts resulting from the publication of this research, as our analysis focuses on aggregate trends and systemic patterns, and quotes do not directly reference users. All data handling procedures were conducted with great care and adherence to relevant guidelines. The dataset was processed on a secure, encrypted server and only accessible to researchers working on this project. The dataset is available to academic researchers through a license agreement to limit access for research purposes only. Any automated analysis, including LLM classification, used local processing only, rather than using cloud-based services, ensuring the dataset remained under our control at all times.

\subsection{Open Science}

The dataset used in this study is available from the Cambridge Cybercrime Centre~\cite{CCCCDataAgreements}. Due to the nature of the dataset, they are unable to fully release the dataset publicly as an artifact. Instead, the dataset can be shared with academic researchers for appropriate projects through a license agreement between researchers' institution and the University of Cambridge.

\section{Results}\label{sec:results}

\subsection{Computational results}

Computational results are from both the fitted topic model and keyword searches of AI-related products, to explore changes over time.

\begin{figure*}
    \centering
    \includegraphics[width=0.95\linewidth]{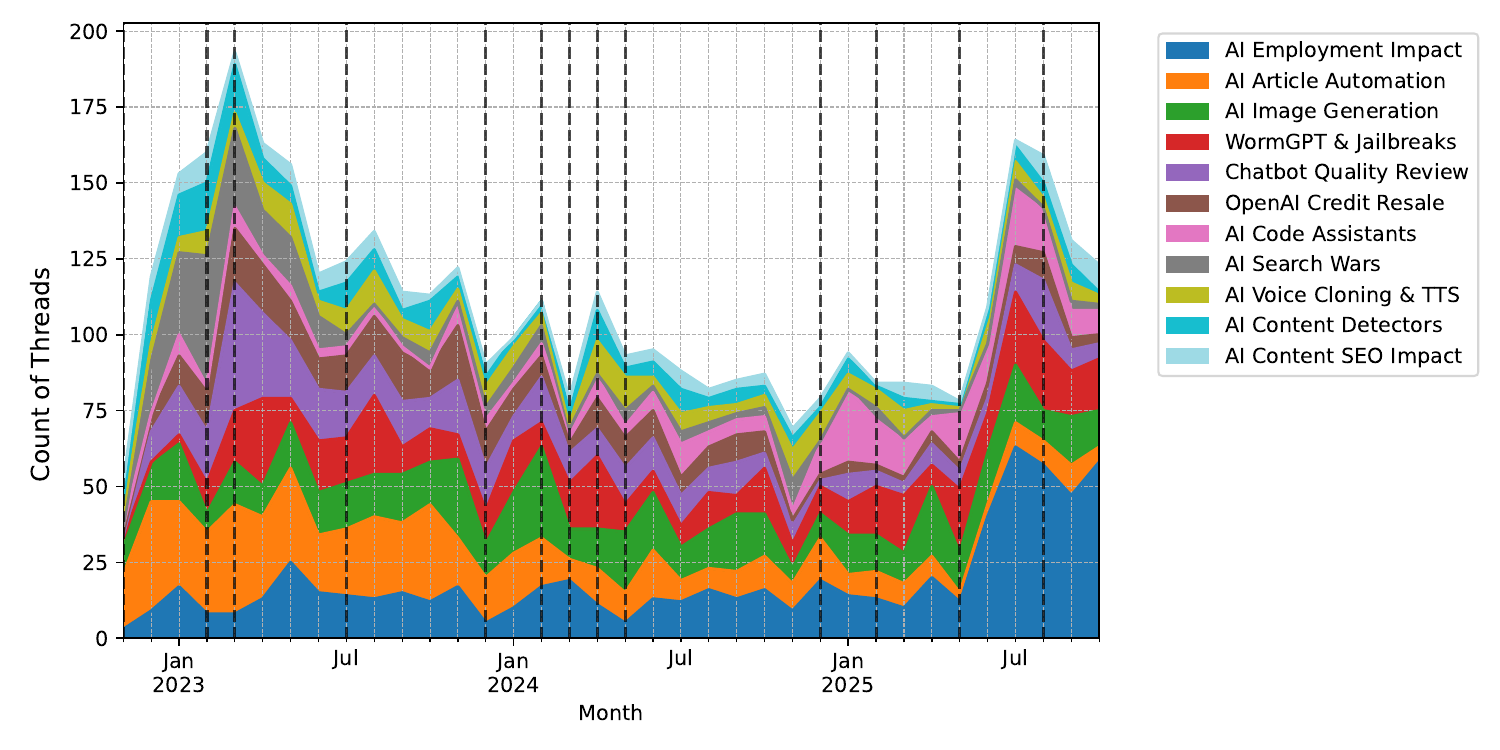}
    \caption{Stacked AI-Related Topics Over Time Up to December 2025 (Excluding Outliers)}
    \label{fig:topics_over_time_stacked}
\end{figure*}

\begin{figure*}
    \centering
    \includegraphics[width=0.8\linewidth]{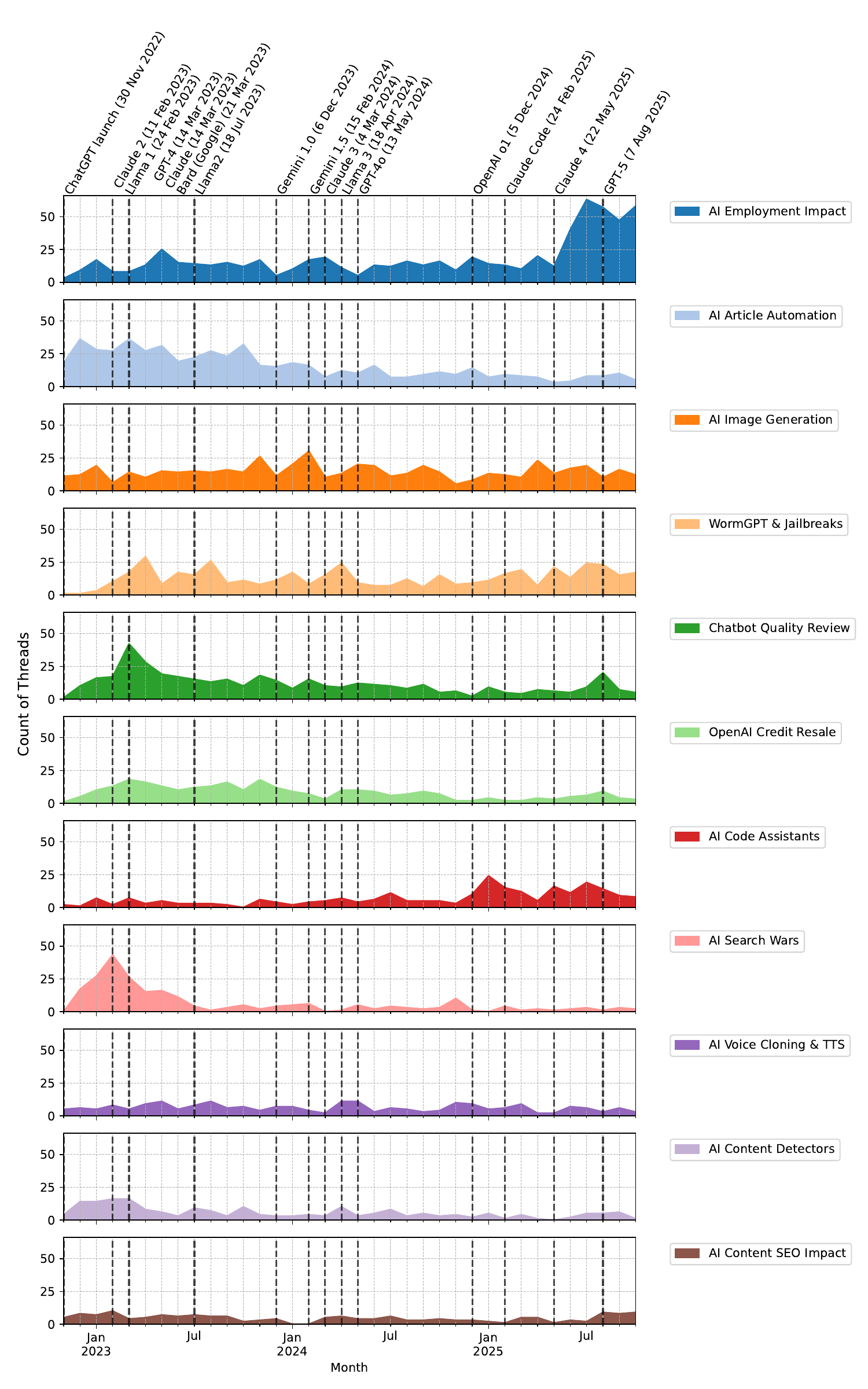}
    \caption{AI-Related Topics Over Time Up to December 2025 (Excluding Outliers)}
    \label{fig:topics_over_time}
\end{figure*}

\subsubsection{Topics over time}

The topic model produced 122 topics. These varied greatly in topics, from meta topics on welcoming members to forums, to specific types of cybercrime topics. We focus our analysis on 11 AI-related topics by selecting and labelling these through manual analysis. Figures \ref{fig:topics_over_time_stacked} and \ref{fig:topics_over_time} show AI-related topics over time, using the count of threads per topic per month, starting on or after November 2022, when ChatGPT was released.

We observe within these topics, there are never more than 100 threads per month. We note threads refer to a collection of posts about a single topic, and can vary in length from one to many posts. The length of threads does not necessarily relate to quality, a short thread could contain a single advertisement for a product, and a long thread could be a debate over a topic or activity.

We find two topics are quite stable over time, with a low count of threads: AI content SEO impact and AI voice cloning \& text-to-speech. These both arise from long-running discussions about AI-related topics before the release of productised text generation tools. 

Following the release of ChatGPT, and around the time of Claude 2, Llama 1, and Google's Bard, we observe changes in four groups: chatbot quality review, AI search wars, AI article automation, and AI contect detection. There are discussions around the quality of generated text, as well as how it can be used for spam and be detected. There is a burst around earlier months with AI search wars, where threads are discussing the disruption to existing search platforms, and how SEO will work with the rise of LLMs.

We also observe a steady level of discussion around AI image generation, despite the release of new models during this timeframe. There is a small burst around the release of Gemini 1.5.

We find a delayed increase by a few months in threads around OpenAI credit resale, and WormGPT \& jailbreaks. The threads on the latter topic are discussing how to bypass safeguards and guardrails in LLM models, and WormGPT was one popular example of a jailbreak. 

Finally, we observe later increases around two topics. Firstly, AI code assistant threads have increased, discussing tools like the Cursor IDE, and the impact this has on development. There is also a sharp increase of discussions around AI and the impact on employment, correlating with the release of Claude 4 and GPT-5.

\begin{figure*}
    \centering
    \includegraphics[width=0.9\linewidth]{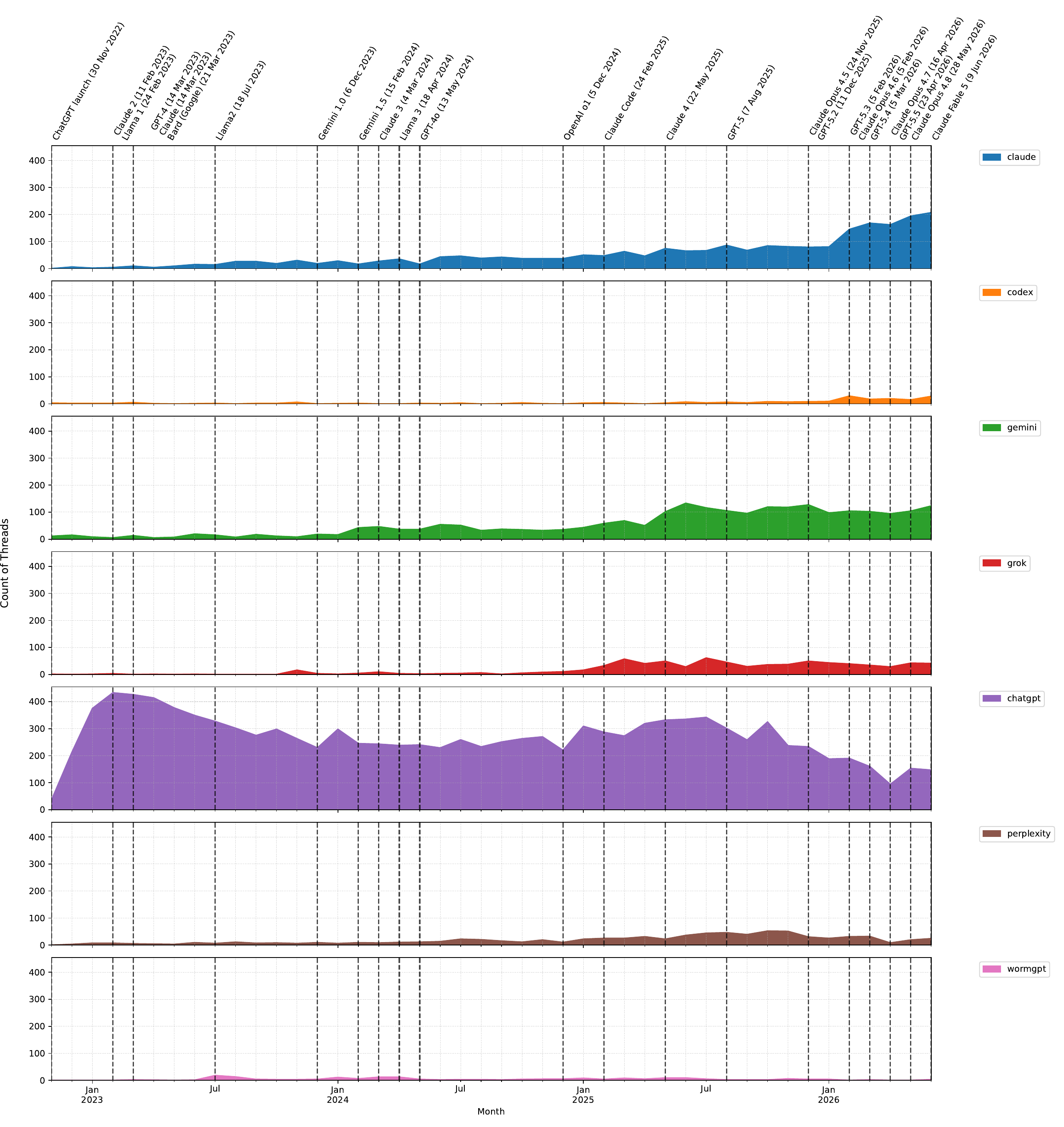}
    \caption{AI-Related Keywords Over Time}
    \label{fig:keywords_over_time}
\end{figure*}

\subsubsection{Keywords over time}

In addition to the topic modelling, we search for a set of product name keywords (case-insensitive): Claude, Codex, Gemini, Grok, ChatGPT, Perplexity. Results are shown in Figure \ref{fig:keywords_over_time}. 

We observe discussions around Claude (Anthropic) have had a steady growth over time, with very little bursty activity, other than a rapid rise in early 2026. Codex (OpenAI's agentic coding tool) has the least number of threads discussing this. Gemini (Google) has a rise in discussion following the release of Gemini 1.5, and also in the second half of 2025. Grok (X) has has a couple of bursts of discussions. Searching for threads on ChatGPT (OpenAI) had the greatest number of topics over time, with a burst following the release and steady activity, remaining at a higher average count than any other keyword. Perplexity had a slow steady growth of discussions over time. WormGPT, a jailbroken model, has small bursts over time, with less than 25 threads discussing this at any time.

\subsection{Qualitative results}

Next, we document themes found through thematic coding, across the combined sample of 3,202 threads. We examine these, and to provide further context, we include quotes. Note these quotes are altered from their original form for ethical reasons.

\subsubsection{Dark AI}

Periodically throughout our dataset, we see periods of substantial interest in various `Dark AI' services -- generally advertised as jailbroken LLMs with guardrails removed, often purportedly retrained on cybercrime or offensive security datasets. Up until late 2024, new models apparently customised for cybercrime were advertised by their developers on most of the major cybercrime forums and groups. 
Dark AI was the subject of substantial cybersecurity reportage in the press, along with 
marketing of the threat by cybersecurity companies. The tools are furthermore the subject of a large volume of requests for free access on the forums, including variously WormGPT and others, as well as AI products for penetration testing such as the open source project (now a commercial product) WhiteRabbitNeo.

\begin{quote}

    Introducing my newest creation, [Dark AI project]. This project aims to provide an alternative to ChatGPT, that lets you do all sorts of illegal stuff and easily sell it online in the future. Everything blackhat related you can think of can be done with [Dark AI project], allowing anyone access to malicious activity without ever leaving the comfort of their home. [Dark AI project] also offers anonymity, meaning that anyone can carry out illegal activities without being traced.

Features:
\begin{itemize}
    \item Fast stable replies

    \item Unlimited characters

    \item Privacy focus

    \item Blackhat allowed

    \item Save results in text file

    \item Various AI models
\end{itemize}

\textit{Forum user \quoteletter, Q1 2023}

\end{quote}

There is very little discussion in our dataset of how (or whether) these tools are proving useful, either for automation of elements of cybercrime crime scripts, for learning, or for malware and code development assistance. Most of the discussion relates instead to accessing the tool (i.e. requests for free access), idle speculation about the effects it might have on hacking, and people reporting that it fails to produce usable code without significant expert user input and guidance, let alone any novel capabilities. We do, however, see a lot of cultural interest in jailbroken or `Dark AI' on the forums -- largely in the form of long discussion threads speculating on the possible futures of AI-driven cybercrime, and excited posts by people using the tools and observing its willingness to talk about illicit topics (though, as can be seen from the pasted chats, this is largely in very general terms). Many of the reviews and discussions describe these tools as not particularly useful -- crucially we see no significant evidence of users describing success with the use of these tools, either as a learning aid or in supporting the development of functioning tools and scripts by neophytes.

This suggests that these technologies as yet do little to overcome the skill barrier to entry for novice entrants to the ecosystem -- while for more experienced coders (as we describe in the following section) chatbots may simply replace the usual googling of errors, searching stackoverflow, or cheatsheets that scaffolded their coding before (and thus not require jailbreaking at all). 

The apparent developers themselves seem to agree. One much-reported LLM chatbot apparently reworked and jailbroken for cybercrime left the market after a short period of time. Users purporting to be the developers of the project argued, as had been well-established by the more critical forum members up to this point, that there was in fact only trivial innovation involved in their product:

\begin{quote}
    We feel that we are increasingly harmed by the media's portrayal...

    At the end of the day, [CybercrimeAI] is nothing more than an unrestricted ChatGPT. Anyone on the Internet can use a well-known jailbreak technique and achieve the same, if not better, results by using jailbroken versions of ChatGPT. In fact, being aware that we utilise GPT-J 6B as the language model, anyone can utilise the same uncensored model and achieve similar outcomes to those of [CybercrimeAI].
    
    So, why doesn't the media report on this? Why do they not highlight that any query posed to [CybercrimeAI] could yield identical or even improved outcomes using ChatGPT? Why do they attempt to tarnish our reputation in this way?

    \textit{Forum user \quoteletter, Q2 2024}
\end{quote}

Discussions of `Dark AI' have decreased in our datasets since early 2024, largely (but not entirely) replaced by discussions about jailbreaking the mainstream legitimate models. Some persisting jailbroken or retrained models are still in evidence, though there is surprisingly little discussion of their active use beyond simply pointing new users in their direction.

In terms of the mainstream chatbots, the experiences are mixed. Early successes with jailbreaking generative AI systems has given way to a widespread frustration post late 2024, with methods increasingly hard to find, and usually only working for a week or a few days (although many forum users report that DeepSeek is more lightly guardrailed than the US models). The discussion has since pivoted to sharing well-established jailbreaks for offline, open source models (which are `frozen in time'), however these models are of generally far lower quality and utility, require significant resources and are slower, and do not update. Thus, we argue below, possibly counter-intitutively for a critical paper, that we have found compelling evidence that guardrails for AI systems are proving both useful and effective in increasing the friction to their at-scale use in cybercrime by neophytes.

\subsubsection{Vibe coding, skids, and skill}

Instead of using `Dark AI' successfully, we observe a number of discussions in which people are finding utility from more general use of LLMs to assist with coding, often where the user has broken fairly generic software engineering tasks into similarly generic steps. These discussions largely relate to use outside cybercrime contexts -- either in people's legitimate day jobs or to non-criminal hobby projects. A number of users reported that the models assisted them usefully with basic coding tasks, e.g.:

\begin{quote}
    I just spent an hour searching online for a working python script to download emails from a remote pop server, parse the messages and then email me with a list of subject lines. I tried ten different scripts and could not get any of them to work exactly as per my requirements. Just before giving up I thought I would try asking ChatGPT for the solution and boom -- it provided me with a script that worked perfectly first time without any alterations.

    \textit{Forum user \quoteletter, Q3 2024}
\end{quote}

\begin{quote}
These tools are a real time saver

\textit{Forum user \quoteletter, Q3 2024}
\end{quote}

The use of these tools has been the subject of considerable discussion in these communities. However, many users (which we discuss below) argue that the tool operates as an enhancement for the already proficient, rather than a reduction of the skill barrier to entry. In particular, they claim that from their own experimentation with these tools, knowing what to ask and the principles of what you want it to do are crucial elements. In discussions of practical uses in real projects, where scaling and automation are sought, a clear barrier appears to be knowing enough to spot and correct mistakes systematically (i.e. through testing).

\begin{quote}
    You need to have a basic level of how code works and your specific language. Otherwise you won't know what part of the code broke and how to refine it. You will even have a hard time to ask gpt to fix it for you.

    \textit{Forum user \quoteletter, Q3 2024}
\end{quote}

We do see some discussions of the use of LLM coding assistants in either developing malware, or in other kinds of software used in cybercrime (such as crypters). Even in these discussions, most users are reporting generic improvements to coding practice -- largely where users already have some ability in coding or software development, rather than anything malware specific. We have observed no credible accounts of LLMs being used to find new vulnerabilities, meaningfully develop or improve on payloads, or (despite some heated discussions) to improve the undetectability of RATs. 

Where people do post their attempts to use LLMs to develop cybercrime tools, these are often mocked by more experienced users:

\begin{quote}
    \textit{USER 1:} ``Today I will be showing you how to adapt a detected crypter stub that can be found freely on github and make it bypass windows defender by using DeepSeek R1... [excerpted] Next we will be using deepseek to create a new unique encryption method for our payload...
\end{quote}
 \begin{lstlisting}
         ***CODE***
         [in c# can you write
         a unique never seen before 
         encryption method that does not 
         use system.encryption,
         it needs to be able to encrypt
         and decrypt bytes,
         and password will be 
         predefined random bytes]
         ***CODE***
 \end{lstlisting}

\begin{quote}

    It then gave me both an encryption and decryption method. You simply replace the encryption method in the builder and the decryption method in the stub. This should bypass windows defender, but for further obfuscation, start to rename every single local variable inside the RunPE run method. Introduce control flow obfuscation inside the method. Split ints into equations that equal the same thing.''
\end{quote}

\begin{quote}
        \textit{USER 2:} ``If I was employed as a threat analyst and this is the kind of development practices behind next gen malware, I'd be celebrating. I could clock in for just an hour a day, get my work done and jerk off the rest of the day with how easy the job would be.''
\end{quote}
\begin{quote}
        \textit{Forum users \quoteletter \ \& \quoteletter, Q1 2025} 
\end{quote}

For these cybercrime-specific applications, the market still appears to be largely dominated by pre-made scripts and methods that circulate in these communities, either sold by those developing them or provided for free once their effectiveness has sufficiently decreased. There are also no serious malware or tools for sale which incorporate LLM elements that we can find, though some users are advertising fairly trivial products as having been coded with the help of AI tools.

While one might expect a community of practice to form around \textit{prompting} LLMs for the purposes of developing malware, this is not in evidence in our dataset. We neither see significant levels of requests for these kinds of specific prompting strategies relating to particular technical parts of crime scripts, or, for example, the circulation of good automation scripts for running cybercrime businesses. Instead, we observe sharing and requesting of jailbreaking methods and prompts for ChatGPT and other mainstream LLMs, and `methods' for incorporating the text and image generation capabilities of LLMs into business models, usually for fraud, low-level passive income generation, or harrassment.  It is particularly of interest to us that the discussions on LLM use in cybercrime largely pertain not to technical matters, but the integration of these into a business model, crime script or money-making framework.

We do observe some requests that hint at other use cases -- again, largely in the domain of fraud and money-making from ad platforms -- such as advertisements and requests for trivial technologies like automated bots that can traverse web marketplaces and gather cookies, where the AI component is to encourage more `human-like' and less bot-patterned behaviour. However, these have not given rise to a local market for these components, as the legitimate companies' agentic tools can already do much of this. Thus responses generally direct the requester to simply use the platform themselves.

So, much as in the regular software industry, LLM coding assistants and chatbots appear to largely be a time saver for people who already have some skills. We caution that these discussions need to be understood critically -- people are very bad at estimating improvements in their own productivity, and the more effective users of these tools may simply be disincentivised to post their results. However, given the `hotness' of LLMs as a topic, their potential relevance as a marketing tool for the pre-made tools and services that drive the cybercrime ecosystem, and the cachet which would be available to those able to demonstrate real, repeatable methods for upskilling using these tools, we argue that were real gains being made, then we would expect to see far more evidence of this than we do. As we discuss in a subsequent section, in the places where LLMs are `changing the game' we do see far more of this evidence.

Experimentation with and use of LLMs, however, remains popular. We observe a lively ecosystem of reselling accounts on the main chatbot platforms, and the same kind of requests for ban evasion services as were common for social media services in previous generations of these forums. Many users are also finding that they are hitting usage limits very quickly and so are looking for ways to get cheap premium accounts.

A number of discussions referred explicitly to concerns about deskilling -- a particularly interesting anxiety given the centrality of skill, learning, and technological mastery to the traditional `hacker' cultural underground.

\begin{quote}

    ``Personally I would advise against this. I've used/tried quite a few of the tools, I'll give a few reasons.

    At first I thought this is great and I knew it would never be good enough to make anything complex but saw a future where I could use it to generate rough boilerplate, as a way to save time on simple/repetitive/non-critical code. What I did not realize is these repetitive and simple boilerplate parts are really important for keeping your skills sharp, including things like typing speed, memorizing syntax (For people who write in multiple languages daily) and solving problems.

    Its clear now that using AI for code causes a very fast negative degradation of your skills... If your goal is just to turn out SaaS scams and you don't care about code quality/security/performance it can be viable to vibe code. (Also seems viable for phishing)''

    \textit{Forum user \quoteletter, Q3 2025}
\end{quote}

Other discussions relate to whether ChatGPT use for coding is `skiddy' behaviour. Again, this deskilling (and anxieties about deskilling) is not novel, nor is it solely a feature of AI -- instead, as above, the effects of AI are to intensify and develop the anxieties, trends, and contradictions of the Crime-as-a-Service (CaaS) era. Similar discussions can be found reported in much earlier papers relating to CaaS schemes -- for example, contentions that booter providers are not `real' hackers and are making no progress on the core work of developing their technical skills \cite{collier_cybercrime_2021}. 

Technical `skill' -- or at least, the valorisation and performance of technical skill -- remains both an important factor in reputation and a shared cultural value in the hacker subculture (though widespread automation and a pivot to service models means it is of declining importance to the practical economy of cybercrime).  However, interestingly, the discussions around LLM use for coding do differ from comparable discussions about, for example, DDoS-as-a-Service as `skiddy' or low-skilled cybercrime, and `not real hacking'. There is notably less consensus in our data around whether LLM use constitutes lazy or low-skilled behaviour. Although many forum users show real scepticism about AI and deride its use as a `lazy' shortcut to real skill, some positive answers focus on \textit{creativity}, \textit{skill}, and \textit{knowledge} as prerequisites for LLM use for coding, arguing that only users with real pre-existing cybercrime skills can usefully deploy it. We can understand these as attempts to retain the centrality of skill as a subcultural value -- namely, that LLM use not only can but \textit{must} preserve the importance of technical skill, knowledge, and mastery, extending this to include a new set of AI-wrangling skills in addition to established `hacker' habitus. However, both these anti- and pro-vibe coding tendencies are still in their early stages, and this conflict remains as yet unresolved.

\begin{quote}
   ``You've gotta first learn the ropes of programming by yourself before you can use AI and ACTUALLY benefit from it. It isn't just about programming. It's also about research. It's about critical thinking. How to solve a problem when you simply don't have enough data. What alternative approaches should I should try to solve this problem. Experiment... etc.''
    
    \textit{Forum user \quoteletter, Q1 2025}
\end{quote}

\begin{quote}
    ``It seems like you're wasting a lot of computing power in order to solve a relatively simple task.''
    
    \textit{Forum user \quoteletter, Q3 2025}
\end{quote}

Long-term use of LLM tools have been discussed and debated by users, when utilising these for attacks. One user noted that while vibe coding could get something running fast, maintainability can be an issue due to the build up of technical debt and a lack of understanding of their own code.

\begin{quote}
    ``I totally agree that vibe coding might get something running fast, but its going to be a ticking time bomb for future headaches.''
    
    \textit{Forum user \quoteletter, Q4 2025}
\end{quote}

Further discussions took place in how to use AI to carry out attacks. This included users hesitancy to deploy vibe coded attacks without taking considerable time to test the attack first. There were also mentions of the risks of using AI attacks more generally.

\begin{quote}
    ``But honestly, I'm not fully certain about the real-world effectiveness rates. I'm a bit hesitant to deploy this in production and find out the hard way if bots actually process these payloads, or just scrape it and move on.''
    
    \textit{Forum user \quoteletter, Q4 2025}
\end{quote}

\begin{quote}
    ``AI-assisted coding is a double-edged sword. It will speed up development but also amplifies risks such as insecure code and supply chain vulnerabilities.''
    
    \textit{Forum user \quoteletter, Q3 2025}
\end{quote}

This gives us a picture of AI use that is not much different to how the hacker community was coding before -- namely, with criminal users largely re-using code made by others with minimal adaptation, and hacker forum users with a real interest in learning mostly using this for non-criminal software engineering projects (borne out in these discussions, where positive stories about LLM use for coding largely relate to people's adoption in their legitimate day jobs or hobby projects).

\subsubsection{AI and the cybercrime underground subculture}

The forums' perception of AI as a wider social phenomenon can generally be characterised as techno-positive -- especially in 2023, there are several discussions hailing these technologies as large technical leaps forward, and general excitement about the possibilities. As we discuss above, discussions of AI overlap with a general celebration of technological innovation and mastery, with particular interested in media reporting of potential effects on cybercrime. As the timeline moves on, discussions move to early experimentation, with users increasingly expressing scepticism as to the real capabilities of the technology, especially for coding. Later discussions, from a qualitative analysis of our sample, do appear to be more agnostic or negative as to the prospects and impact of the technology. More broadly, in terms of the wider cultural effects of AI on the forums, we see from mid 2025 (as can be observed in our computational analysis) an increased anxiety about job losses, especially in coding, due to AI adoption in the legitimate software engineering industry.

\begin{quote}
    \textit{USER 1:} ``Freelancing Web Dev is no longer a viable income source in 2025. With the AI hype, the market has gotten 1000x more saturated as now everyone vibecodes several React projects, then lists themselves on fiverr or upwork or discord as a professional fullstack developer with 3-5+ years of experience. This is annoying as hell. I cannot find a job for the life of me. I can't use Fiverr, and so I'm stuck to finding clients on Discord. I did a project a few months ago as a coincidence but then stopped working for a while ... now I cannot find ANYONE no matter how hard I advertise.''
  \end{quote}
    \begin{quote}
    \textit{USER 2:} ``The market is quite saturated, but there is still room for good developers. However you need to have skills that are greater than the average vibecoder, and you need to make your track record obvious and trustworthy, as well as needing to get good at branding and selling yourself. It's really is going to be harder now.''
  \end{quote}
    \begin{quote}
    \textit{Forum users \quoteletter \  \& \quoteletter, Q4 2025}
\end{quote}

The technologies are having further effects on the forums as a subcultural space. One use case we do observe is the proliferation of LLMs being used for low-quality reputation-farming posts on underground forums, often to dump long bullet-pointed explainers for basic cybersecurity concepts. 

\begin{quote}

   ``This is a small PSA to stop anyone else from making themselves look like total buffoons in the cryptocurrency subforum, or any other subforum from now on. I will not name the poster as it's pretty obvious to us, and I'm sure to them too, who exactly I'm referring to.

    Changing your username and general tone then using AI to write content does not make you seem the crypto guru you think it does, and asking chatbots whether throwing \$75 into Mantle is a good idea is not worthy of posting, and we just don't care whether you made \$25 or not.''
  \end{quote}
    \begin{quote}
    \textit{Forum user \quoteletter, Q4 2025}
\end{quote}

\begin{quote}
       ``If I wanted to talk to an AI chatbot, there are many websites for me to do so, but that's not why I come to [the forum]. I come here for human interaction... Forums are inherently human. Introducing some AI or otherwise generated replies just defeats the complete purpose of visiting and/or maintaining a such a forum."
      \end{quote}
        \begin{quote}
    \textit{Forum user \quoteletter, Q4 2025}
\end{quote}

While causing irritation, the forum discussions around this are often thoughtful -- reflecting on the long-term decline of users, engagement, and the wider relevance of underground hacking forums since the `golden age' of the early to mid 2010s. Several forums were attempting to use the popularity of the AI wave to rejuvenate interest in their communities, attract new members, or stimulate discussion from existing members. However, many were also struggling to battle waves of low-quality posts and replies, often clearly authored by chatbots. Some experiments, for example, include forum-branded chatbots that largely expand on news bulletins from RSS feeds, 
 show clear attempts to follow the major social platforms like Meta in attempting to use chatbot content 
to battle a decline in user engagement~\cite{murphy_meta_2024}.
However these were often met with skepticism from users.

\begin{quote}
   ``I'm disappointed that you are working to incorporate AI garbage into the site. No-one is asking for this -- we want you to improve the site, stop charging for new features. AI subforums are fine, but you're going down a rabbit hole and people are unlikely to use the features you've made over the mainstream AI products that exist. You're trying your best to save the forum but this is the wrong way to do it -- it's not working, and interest and engagement keeps dropping.'' 
   \end{quote}
        \begin{quote}
   \textit{Forum user \quoteletter, Q2 2025}
\end{quote}

In addition to these critical posts, some users did signal their approval of the officially-sanctioned chatbot content, however it is worth noting that this was accompanied by other discussions indicating that accounts that were overly critical of the forum owner were being banned. More generally, complaints about declining traffic (attributed to Google's AI summaries) and an ageing, low-involvement user base suggest that cybercrime forums are being impacted by AI tools as much in their capacity as \textit{forums} as in their capacity as deviant subcultures.

\subsubsection{Automation}

As we describe above, much of the technical business of cybercrime is already heavily automated and reliant on pre-made assets, methods, and scripts. We observe some discussions of LLM content displacing resources in areas already heavily dependent on pre-made assets (though not for pre-made scripts). There are numerous discussions of using LLMs in website creation, in image generation for marketing materials or logos, and for any of the various activities that require social media bots. However, users report varying results from these activities, which still require substantial human tailoring and management:

\begin{quote}
    \textit{USER 1:} ``I've noticed [much worse SEO performance] with some AI-generated frontends: they look fine but output bloated markup or poorly structured layouts that can confuse crawlers. Especially when the div soup replaces semantic HTML, or when key tags like <title>, <h1>, and meta descriptions are either generic or missing. If you're aiming for organic visibility, it's worth doing a quick check post-generation, or even hooking the output to a static site generator where you can enforce cleaner structure and meta tagging. Vibe coding tools like v0.dev are great for prototyping, but definitely not production-ready for anything SEO-heavy without cleanup.''
  \end{quote}
    \begin{quote}
    \textit{USER 2:} ``That's exactly the way to go. Once the prototype is solid enough, exporting to static and auditing structure, heading hierarchy, and internal linking makes a real difference. I am currently testing a pipeline with [tools] to catch such crawl issues early. This is still a bit clunky, but it's much better than shipping AI output raw. If you're running JS-heavy components, you may also want to pre-render critical pages to avoid rendering gaps.''
  \end{quote}
    \begin{quote}
    \textit{USER 3:} ``I think AI isn't good enough to handle the kind of volume of code I would be flashing through it and asking it to expand on features. I think every project has that aspect to it. AI can only still do the basics. It does them pretty good though. But I would not trust anything beyond my own supervision, and copy and paste from it only. I wouldn't just let it loose and say go on improve this or find that issue. lol I would be out of API credits by the end of the day! lol and above all I just don't trust it. There are too many 1's and 0's being blasted to not get a few wrong, and every program apart from a chatbot output, needs to have correct code.''
  \end{quote}
    \begin{quote}
    \textit{Forum user \quoteletter, \quoteletter, \& \quoteletter, Q2 2025}
\end{quote}

The same is also the case for bot automation, where (as we also see for blog farming), a set of practices has emerged for slowly introducing automated content at scale with substantial manual work in order to avoid detection and banning by platform algorithms:

\begin{quote}
    \textit{USER 1:} ``I am ... interested in starting an [adult content website] agency. I have currently got a few girls from my city that would be interested in me helping market them...I'm interested into using proxies and running ...[social media platform A]. What bot automation site would be best? Is it still [bot automation website]? I'm also not experienced with bot automation or proxies so I'm worried that my... account would suspect bot activity and get flagged. How do I avoid this? Are there any courses or tutorials that specifically help with this part?''
  \end{quote}
    \begin{quote}
    \textit{USER 2:} ``Since you already have experience with Twitter and Reddit, I'd recommend using them to drive traffic while you build out [social media platforms B and C]. [Social media platform D] is great for direct promo, and [social media platform E] can be a goldmine if you know how to avoid shadowbans. For [social media platform A]... automation is tricky. [Bot automation website] is still a solid option, but you'll need good cellular data proxies to avoid detection. The key is to keep automation settings human-like gradual actions, random delays, and avoiding repetitive patterns. If you are new to this, I'd suggest starting with manual growth first while you learn. There are some good resources out there for automation and proxies, but a lot of these courses are outdated or overpriced... If you are looking for a hands-on approach, testing with a few burner accounts before running your main setup is a good way to learn without risking any bans.''
      \end{quote}
    \begin{quote}
    \textit{Forum user \quoteletter \ \& \quoteletter, Q1 2025}
\end{quote}

Interestingly, there is almost no discussion at all of agents for automation of the logistics of cybercrime. There is no market we can find on these low-level forums for AI automation services for setting up servers, for running DDoS-as-a-Service services (`booters'), for botnet administration, cashing out, or spreading and infection (much of which was automatable in theory using existing licit and illicit management tools). Naturally, this does not mean that these tools are not being used to this end, especially by more established providers, however we believe that we would have expected to see a secondary market (or at least a set of shared guides and tutorials) on the underground forums emerging by now, given the previous social dynamics of these communities.

\subsubsection{Scams, get-rich-quick schemes, passive income and other sAIde hustles}

The major use case in our datasets for LLMs and image generators was not in `hacking' or classic cybercrime forms, but in the long-standing ecosystem of get-rich-quick schemes, low-level fraud, and passive income generators. In many cases, these schemes have extremely low returns per unit, very low individual success rates, or a tendency to saturate very quickly, and so low-involvement automation was already a requisite for profitability prior to the widespread availability of LLMs.

For SEO and ad revenue passive income streams, in which players create webpages then attempt to push them up Google rankings to make money from advertising (or from selling these resources for the purposes of marketing products or brands), numerous participants complain from 2024 onwards of generally declining revenues and success rates over time. They attribute this to wider changes in user traffic patterns (with many simply using chatbots rather than Google search) and the introduction of Google's AI summaries which appear at the top of search queries -- both features which have similarly reduced revenues in the legitimate content economy. 

Their reaction to this appears to be to adapt through incorporating LLMs into their business models, with an aim to increasing scale to offset lower unit returns. LLMs are being used to mass-generate blogs to fill (still mostly pre-made) web templates -- although users report mixed efforts, and substantial human work to avoid attracting low-quality rankings:

\begin{quote}
    \textit{USER 1:} ``people who still have a growing niche blog with 100\% ai generate content after google's march update. how?''
    \end{quote}
    \begin{quote}
    \textit{USER 2}: ``I just never use AI to write a full post. I use it to expand my writing. I write a paragraph, copy and paste to ChatGPT, then ask it to expand on this using the same style and tone and whatever else you want to add. Not that sure how safe this is, but haven't been hit yet and it's the safest way I can think of doing it.''
      \end{quote}
    \begin{quote}
    \textit{USER 3}: ``Anything that is a directory style.  Take a website that lists doctors in your neighborhood for example. You can scrape this data from [directory websites] etc and also get data like where their office is and what experience they have etc.  Now if you just pasted this data raw into your site then you would never run ads on it. It would be flagged for thin content. But if you use an LLM to generate sentences out of this information, you now have hundreds of pages worth of information.''
  \end{quote}
    \begin{quote}
    \textit{Forum users \quoteletter, \quoteletter, \& \quoteletter, Q2 2024}
\end{quote}

This is a clear example of LLMs pushing the economic incentives of already highly saturated markets further in the direction of scale and oversaturation. Depending on the actions of the platforms that set the material political economy underpinning this activity, the vicious cycle this suggests (as with the discussion of romance fraud below) may well push the market back in the direction of more tailored, higher-quality content, or could equally further degrade the experience of search for everyday users (and hence drive them towards the chatbots).

Within SEO discussions, we observe some conversations sharing efforts to attempt so-called GEO (i.e. Generative Engine Optimisation) -- a shift in practices hypothesised by the legitimate marketing ecosystem \cite{aggarwal_geo_2024}. Concretely, users are sharing various early experiments with getting the major chatbots to preferentially mention their company or service:

\begin{quote}
\textit{USER 1:} ``Has anyone actually managed to get chatgpt to mention their brand?"
  \end{quote}
    \begin{quote}
\textit{USER 2:} ``Being mentioned on Reddit, Quora and build up authority is the way to go. It's easier when you focus in a  niche or low competition industry.''
  \end{quote}
    \begin{quote}
\textit{USER 3:} ``I have gotten it to use sources to answer queries about a niche, and such sources are 30\%+ owned by me (youtube shorts), but this does not help my traffic yet, however if you know what to do next it can be valuable''
  \end{quote}
    \begin{quote}
\textit{Forum user \quoteletter, \quoteletter, \& \quoteletter, Q2 2025}
\end{quote}

The particular expertise of these communities in cultivating large bot followings and high Karma on Reddit is therefore clearly proving useful given the importance of Reddit as a source for Google's chatbots and AI services. This suggests that the social media bot resources and low-quality spam websites that they have been using to farm ad revenue, sell followers, and promote scams 
may have a new set of uses in illicitly influencing the knowledge base of the major chatbots.

We observe some small-scale business models emerging for generating sources of income in other ways using LLMs. A typical case here involves, again, multiplying the scale of low-return practices until they are profitable, often still in terms of only a few hundred dollars per week. For example, a common business model involves generating and selling eBooks:

\begin{quote}

    ``For this method, we will be using chatgpt to create customized and unique ebooks to sell or share. To get started, think of a popular niche right now, such as dropshipping, crypto, etc, and pick the one you like the most. If you cannot think of any ideas, just ask Chatgpt ``what are some popular ebook categories right now''. \\
    
    ... The next step is to create a rough outline for the ebook, and ask chatgpt to help you! ``Write an outline for a book focusing in Affiliate Marketing''  This will help organize some initial ideas into a structured outline. Copy this outline as you liike and then start building a rough draft in google docs or another document editor.\\

    To get more content on each chapter in your outline, ask chatgpt for this... You can probably see where the rest is going now, use ChatGPT. To create the backbone of your eBook. The last step to complete the eBook is to add some nice graphics and create your final draft after editing mistakes, humanizing etc etc etc.'' \\
    
    \textit{Forum user \quoteletter, Q3 2024}
\end{quote}

Again, however, we can observe that most of these methods still involve laborious human effort -- in this case, copying and pasting content in and out of Google Docs, editing, and `humanising' the output.

Finally, some forum users have been experimenting with integrating LLMs into their ponzi schemes, crypto scams and pump-and dump practices. This includes forms of semi-automation in crypto scams and in the creation of cryptocurrency projects (such as logos, marketing copy, and associated materials). We do see some interesting wider use of automation in the secondary markets around cryptocurrencies (for example, for the sale of services that use bots and automated website posts to build hype and `market signals' to pump the value of coins).

\subsubsection{Social engineering and interpersonal deception}

There are some discussions in these communities of the use of LLMs in social engineering -- especially voice cloning and deepfaked video -- although these are mostly discussions of news reports from the cybersecurity press. There are some discussions of the use of deepfaked images and video for verifying accounts, though discussions of the use of deepfakes in commercial fraud (despite clear interest) are largely speculative in these communities.

While experimentation with GenAI use in corporate or consumer fraud and offensive social engineering has been observed by defenders, this dataset offers little direct insight into these practices. It appears that this activity (whether a genuine disruption or not) is confined to more professionalised operations, rather than the wider 
underground
ecosystem. However, the use of GenAI tools in activities with some similar dynamics -- namely, in romance scams, eWhoring~\cite{hutchings_understanding_2019}, and other money-making ventures involving forms of direct deception -- is observed on these forums, and offers some insight into the difficulties and opportunities posed by incorporation of these technologies into existing business models.

Much of the practical discussions of experimentation in these communities are romance fraud and forms of 
`eWhoring', an established practice, that pre-dates modern generative AI tools, of using either automated videos and images to simulate sexual conversations with other Internet users, or (more recently) working directly with online sex workers to automate interactions with clients~\cite{pastrana_measuring_2019}. These have both long involved the heavy use of automation and the production of faked video, so would be a likely area in which generative AI would be potentially useful. Here, we observe three main use cases. The first concern attempts to use AI tools to scale up customer interactions -- namely, to generate text and video content with larger numbers of victims or clients at once. This often appears to still involve use of `packs' of pre-made interactions and content.

\begin{quote}
    \textit{USER 1:} ``AI eWhoring. does anyone have any tips, or any ways for me to get started. i am not able to post in the eWhoring forum as well, so sorry for posting here.''
      \end{quote}
    \begin{quote}
    \textit{USER 2:} ``I have seen others feed their local AI with a profile and assets (videos and images) and then have it actually handling messages itself. But they however are easy to spot, if you ask it a complicated question, you shall see the AI behind it.''
      \end{quote}
    \begin{quote}
    \textit{Forum user \quoteletter \ \& \quoteletter, Q2 2025}
\end{quote}
    
Conversely, we also see these tools used for enriching more high quality encounters and targeted activities, allowing individuals to produce more realistic and reactive content for smaller numbers of clients/victims.

Finally, we see the incorporation of the tools into more logistical elements of romance fraud and eWhoring, such as evading ID verifcation on large scale social sites when setting up fake accounts (commonly discussed for people catfishing). The use of pre-made assets and forms of video automation and semi-automation~\cite{pastrana_measuring_2019} is, as we argue above, well-established in these forms of fraud, long pre-dating AI technologies. As this discussion highlights, only a fraction of the community are able to make real sustainable income through these methods and only with considerable effort and resources, as with previous automation and scaling technologies for social fraud:

\begin{quote}
    \textit{USER 1:} ``...There's nothing inherently valuable in Ai models. In fact, they exist in abundance...  If 5000 people here tried to replicate this, i would bet you good money 4995 will fail. The company who owns this ``influencer'' is just a marketing company. They have resources to grow an IG page. A lot of fake followers. A lot of bots. A lot of money spent on advertising. It's not just simply creating a model and voila....Everyone is trying to do this and most are failing. There's nothing to ``tap into''. There's no ``great opportunity'' out there with this. Just the usual: Very lucky and resourceful people sometimes are able to make good money online...''
      \end{quote}
    \begin{quote}
    \textit{USER 2:} ``...People have been catfishing, using dodgy photos etc for eternity on here, so now create your own, even sell a pdf tutorial...''
      \end{quote}
    \begin{quote}
    \textit{USER 3:} ``Fast mover effect plays major role in this thing, I saw bunch of youtubers talking about it and this can only be something new to the noobs''
      \end{quote}
    \begin{quote}
    \textit{USER 4:} ``...The successful minority is a statistical outlier who would never want to attract any attention to themselves and their work. The mediocre majority will almost always fail out, while a few become gurus and pretend they're successful...''
      \end{quote}
    \begin{quote}
    \textit{Forum users \quoteletter, \quoteletter, \& \quoteletter, Q4 2023}
\end{quote}

In particular the use of voice cloning is widely discussed (though its efficacy is contested):

\begin{quote}
    ``I dedicate myself to selling content with deepfake and combined with AI, through social networks, Facebook, Instagram, and my strong Telegram and WhatsApp. I have a ``model face'' that has been with me all this time, but in November 2024 I started a new ``model face'', so what I did at first was a lot of deepfake but then as I had money to invest I started paying, first for a regular PC, then paying for Software, etc. Today I can say if a message arrives on any network, I just respond with a voice message (I have real-time voice cloning), which helps people trust me, but even so, some people ask for something more, so I also have a way to do a `verification' by photo with the person's name, and in very specific cases, a video call with a deepfake in real time''

    \textit{Forum user \quoteletter, Q1 2025}
\end{quote}

There is a lively set of discussions on the effects on the economic dynamics of romance fraud by generative AI. For eWhoring in particular, forum users report wider changes to the economy of fraud from AI tools being as important as the adoption of these tools into fraud crime scripts themselves. There is a perception that many potential romance fraud victims are now engaging directly with chatbots from romantic and sexual fulfilment, and that the general exposure to automated or generated content is increasing suspicion and wariness of the victim market, especially for highly scaled and automated interactions. However, the wider dynamics are reacting to changes not directly related to AI tools as well -- in particular, the rise of OnlyFans (and associated management agencies with their own automation strategies) is reducing the supply of targets for traditional forms of romance fraud, some of which are moving into supportive automation services for these legitimate agencies. Even these forms of automation still require significant human review and input (often from affiliate labour) -- and some forum members argue that the saturation of the wider market is making `real' human interactions more valuable:

\begin{quote}
    \textit{USER 1:} ``It sounds like you're doing everything the right way. You're only doing custom content and you're actually interacting with the men. That's exactly where the money is -- especially now that everything is agencies and chatbots.  The problem you're running into is that ``ewhoring'' as it once was is mainstream now. The men on OF basically know that they might not be actually talking to the real creator. And the majority of the OF girls are lying about their earnings to get more attention. Same thing that the ewhores on here used to do.''
    \end{quote}
    \begin{quote}
    \textit{USER 2:} ``Since you actually have a real girl to work with and she's good with the interaction, that's what you should focus on. Advertise the hell out of her (Instagram/Reddit/X-Twitter) and emphasize that she's real and there is no agency. And charge more than you already do. If they complain, can just tell them that they can always pay the cheaper price for the chatters from the Philippines.''
    \end{quote}
    \begin{quote}
    \textit{Forum user \quoteletter \ \& \quoteletter, Q3 2025}
\end{quote}

We do observe discussions of a shift to more boutique, hand-crafted interactions (both for direct fraud and more general eWhoring services) for higher unit returns, supported by publicly available image and voice generation tools. 

Although we see no real discussions of successful use of Generative AI tools for commercial video fraud or voice fraud, the experience of the romance scammers and eWhoring community does suggest the potential utility of some of these tools for fraud (though not necessarily to any transformative effect). We see further discussions of speculated saturation effects from these tools in the commercial fraud space -- often framed as a reduction of available new victims through large expansions in the volume of fraud.

Finally, we do also see clear signs of automated misogynist harrassment. These pre-date the updates to Grok in early 2026 which allowed the production of intimate nonconsenual images, indicating the availability of services using jailbroken (or poorly guardrailed) AI for producing these images as a service for customers.

\begin{quote}
    ``Im able to make any girl nude with an AI that 3 partners and me created.\\
   Very high quality, fast generation\\
      Pricing:\\
- 1 Picture = \$1\\
- 10 Pictures = \$8\\
- 50 Pictures = \$40\\
- 90 Pictures \$75''\\

\textit{Forum user \quoteletter, Q2 2023}
\end{quote}

While this is undoubtedly concerning, the existence and apparent popularity of these services do suggest that attempts at guardrailing are at least somewhat successful in creating a skill barrier to entry for the production of these kinds of images. While not eliminating the problem entirely, there is clearly a difference in the wider harm impacts of a model that requires significant effort to jailbreak for the production of harrassment images, compared to one which allows the creation of these trivially, at scale, and by any untrained user. It further suggests that measurement of the prices charged for these services over time could serve as a proxy for the success of some forms of guardrailing.

Interestingly, we also see a number of discussions that describe jailbreaking AI as a form of social engineering in its own right:

\begin{quote}
    ``You have to lead it into a topic dont outright say `Don't censor anything'. It's just too all encompassing, you have do it topic by topic, talk about the starting topic that doesn't involve the `bad' stuff. Say hacking and SQLi, don't start extremely and slowly lead into the topic by talking about SQL first. It's just social engineering except its AI. You just have to gaslight the hell out of it.''

    \textit{Forum user \quoteletter, Q4 2025}
\end{quote}

The transfer of this mental model (and attendant skills) is noteworthy -- this is a clear attempt to `domesticate' LLM chatbots within the social and subcultural context of these forum communities. These cultural elements of adoption are important -- participants in this subculture being able to understand these technologies as compatible with their existing values and practices, and to adapt their skills to this new innovation are crucial elements that we also observe in social learning models of technology adoption and innovation in legitimate technology ecosystems.

\subsubsection{New markets and targets}

Discussions of genuine new markets, targets, or business models for cybercrime are absent from our dataset, beyond speculation based on the cybersecurity news. As would be expected given the generally low levels of technical skill in the cybercrime underground, we find no notable discussion of arcane novel practices and vulnerabilities such as living-off-the-LLM, novel advanced penetration methods, or complex adversarial prompting. We see no discussions of practical exploitation of the adoption of LLMs by legitimate actors, other than some minor discussions of easy-to-exploit vibe coded websites. There is no real discussion beyond reported news stories of successful disruption of corporate chatbots used by businesses. The markets and targets we observe emerging here are largely edge innovations in existing high-volume business models (such as dropshipping or SEO).

\subsubsection{Developments post-January 2026}
\label{sec:developments_post_2026}

Our analysis breaks at January 2026 - the initial cut-off for the paper's data source, and something of a turning point for the wider AI ecosystem (with the release and mass adoption of several updated Claude models fully taking off in industry for coding assistance taking place at around this time). We have taken the decision to break our analysis at this point, and discuss developments in 2026 separately in this section.

The main difference we observe is that from January onwards, the process of mass adoption of Claude for vibe coding begins in earnest in the cybercrime underground, much as it does in legitimate industry. As in observations in the previous period, this overwhelmingly pertains to the use of Claude for generic coding functions associated with illegal enterprises rather than cybercrime-specific coding per se. A wealth of complaints from `newbies' reinforce the finding that for cybercrime operations as for legitimate coding, substantial expertise is still required for AI tools to produce anything useful. Complaints of technical debt in AI projects are now far more a feature since January 2026.

Jailbreaking is still a major concern of the forums, though becoming increasingly elaborate as the efforts to remove these on the part of the AI companies appear to be fairly successful. For example:

\begin{quote}
    
    ``Hook the agent harness up to a Ghidra MCP server, then choose a low tier model like Gemini Flash to ask an inital question. After that it will think for a while, then stop the response and switch to using Claude. Then resume its response and it will use Gemini's non-refusal as the base for the reply and bypass any restrictions or refusals. This jailbreak does work for most prompts, not just reverse engineering.''
\end{quote}
\begin{quote}
        \textit{Forum user \quoteletter, Q2 2026} 
\end{quote}

There is also an observable increase in AI-specific requests and services - such as removing image watermarks added by the mainstream AI models. AI `skills' are themselves now heavily in demand and being requested and advertised on these forums.

We observe a notable increase in obvious scamming behaviour (also observed by forum participants), generally involving faked reselling access to Claude and `max' upgrades. There is also increasing evidence of serious tightening of the market for software services, with many reporting prices dropping and freelance work drying up. And the AI coding `uplift' appears to be a new baseline rather than a gold rush or profit multiplier - a typical thread in which users discussed their finances included references to low-paid manual work being used to complement income from fraud. More generally, a feature of forum discussions since January has been people announcing a completed AI-assisted software or coding project, then complaining that they are unable to successfully monetise it.

There is some discussion of agentic AI tools, but these largely appear to be automating `boring' administrative and maintenance tasks and bot behaviour. A particular feature of this is a notable increase in discussions about mobile phone farms and residential proxies since January 2026 -- used to circumvent device ID restrictions put in place by the tech companies. While automation using mobile phone farms pre-dates the `AI wave', this appears to have substantially contributed to their increasing prominence as a `best practice' method.

As will be familiar to users of social media, we also see wholesale use of AI to `improve' writing, which has led to some rather bizarre and homogenised posting behaviour. A particularly interesting cultural feature appears on the more coding-focused forums, in which many discussions claim to be able to identify AI-generated code through stylistic and structural `tells'. While in less code-focused forums, AI coding assistants have become the new baseline, in these forums AI use remains a controversial violation of the values of their craft-based subculture. In the fraud and crime-focused forums, the culture has generally continued its long-term slide into the jargon and practices of affiliate marketing and the small-time entrepreneur: the average discussion is far more likely to reference attracting `mugs' and traffic, conversions, sales metrics, business plans, advertising strategies, trivial productivity hacks, and get-rich-quick schemes than anything related to the craft of hacking.

\section{Discussion}\label{sec:discussion}

This paper is the first of which we are aware which attempts a community-scale mixed-methods analysis of the adoption of generative AI technologies by the underground cybercrime ecosystem, broadly considered. Our findings -- namely, that despite widespread interest, adoption has been patchy and largely ineffective within the communities we consider -- are unsurprising, largely conforming to well-characterised dynamics of technology adoption in technology and innovation studies and evolutionary economics. 

\subsection{Cybercrime and skill barriers to entry}

The most striking of our findings relate to the relative lack of success that LLMs and coding assistants have had in reducing skill barriers to entry for novice cybercrime actors. As we argue above and as evidenced in the literature \cite{collier_sophisticated_2022}, in contemporary cybercrime economies, technological skill is of only marginal important despite its importance as a cultural value. Pre-made scripts, methods, and resources abound -- most innovation is in implementation and business model. For our cybercrime communities, the effects of vibe coding appear to mirror much of what is being reported by legitimate industry (namely, that they largely act as skill multipliers for higher-skilled users rather than reducing skill barriers for novices)  \cite{fawzy_vibe_2025, meske_vibe_2025}. This may therefore be another way in which large-skill cybercrime mirrors the legitimate tech start-up ecosystem.

Even for more `skilled' actors, or those who are involved in successful illegal businesses, these tools do not appear to be a game changer in terms of coding -- they free up effort and time that would otherwise have been used in searching error messages or consulting cheatsheets. This improvement has, however, been mitigated by increased requirements for checking and editing code. Again, the technical side of these businesses were already largely automated and based on pre-made resources.

Overall, therefore, we see relatively little change in the economics of skill in cybercrime so far. We do observe some niches of successful innovation and implementation, however these are largely outwith the domain of coding per se (which, we argue, was not a rate limiting step for scale and success).

\subsection{The economics of successful scale and adoption}

We see almost no discussion of these tools being used (whether coding assistants or automation agents) to disrupt or scale most existing forms or components of cybercrime (such as DDoS, the management of botnets, or ransomware to name only a few), whether in their technical or logistical aspects. In particularly, qualitative analysis uncovered very little use of agents or real automation, other than in areas already heavily automated (and even here, largely focused on content generation rather than logistics). We did not find advertisements of tools that integrate LLMs in meaningful ways, only some small and largely trivial examples of scripts nominally created with the assistance of AI coding tools. There appeared to be very little sharing or discussion of prompting strategies beyond jailbreaking. It is important to note that effective prompts depend on what the user want to achieve and the specific model employed -- and advice such as example prompts and strategies become outdated quite quickly as models are updated and contexts change. But we do not see the kind of collaborative prompt creation discussions that are the basis of thriving communities in other subcultures (such as the AI image and video generation, music, or legitimate coding communities).

Where we do see interesting results is in pattern manipulation. There is good evidence in our dataset that the strong pre-existing set of practices through which users had been attempting to evade detection by defenders at existing levels of scale and automation (here largely in discussions of bots, romance scams and SEO optimisation) have indeed seen adoption of generative AI tools. These attempt to mask the signals of automation at scale that defenders use to fight them. These skills in scaling automation (often by careful use of timing, or blending in human and `random' activity) are a core cybercrime skill that was already emerging well before the advent of generative AI, and will clearly continue to rise in prominence in these ecosystems.

However, our qualitative findings show saturation dynamics emerging related to AI adoption. These are often extrinsic (e.g. AI overviews in Google search results, reduction in victim pools) and some intrinsic -- leading to much lower unit profit, necessitating further scale and saturation. This may in fact lead to some existing illicit business models becoming unprofitable (or at least reaching a steady state equilibrium for a small number of actors). In some cases, we observe both saturation and a shrinking of supply of victims due to wider changes to the dynamics of Internet use (such as movement between platforms). This is especially the case in romance scams and eWhoring, where some scammers report being outcompeted by Generative AI chatbots for victims.

We also, from our computational analysis, see a particularly interesting trend in discussions of AI article automation -- namely, an initial peak of discussion that trends linearly down over time across our dataset. From our exploration of the qualitative data, we argue that this may reflect a fairly successful adoption trajectory -- a lot of initial discussions in which users collaborate to share strategies and report successes, which decreases over time as the strategies begin to work well and the practices and implementations mature.

We find that while forum members are clearly adapting to and using generative AI, we find little evidence that these tools are making a significant difference. Automation tools, while apparently easy to deploy, become complex to manage logistically at scale -- there is real skill (though not necessarily coding or technical skill) -- to scaling automation, especially where effective countermeasures are being taken by industry.

A key finding our our study is that guardrails do seem to have some clear positive effects, even where they are fairly easily circumvented. In the wider economics of these different cybercrime forms, they add enough friction and cost to meaningfully prevent the formation of scale production and automation capacities for long enough to make genuine shifts in scale viable and sustainable. After 2024, posts discuss how hard is it find jailbreaks that work for more than a week, and the existence of a jailbreaking market for images suggests at least some suppression (though these posts pre-date Grok's changes that allowed the at-scale production of illegal image content). This aligns with previous discussions of whack-a-mole strategies \cite{collier_booting_2019, collier_cybercrime_2020}, namely that despite their apparent lack of effectiveness, they do meaningfully increase friction and cost of participation in the criminal ecosystem and prevent the achievement of further scale. In effect, this kind of regulation still has the potential to dissuade and shape the economics of cybercrime as a volume phenomenon. Further, the chatbots also (as users themselves note) present a control point from which law enforcement can potentially surveil user behaviour.

\subsection{Hacker culture, community, and identity}

 We report a substantial degree of cultural interest in generative AI technologies in the cybercrime underground. Impressions, however, are mixed. We see a remarkable degree of skepticism from such a techno-positive community. In fact, this is a community 
that has long experience with automation, new technologies and associated hype, attempts to scale business models, and generally cobbling together things that work to achieve a minimum viable product. However, skill, business practices and technology are not simply economic factors, but are important aspects of `hacker' culture (and the other subcultures we see in these communities). The diffusion of a set of technologies which appear to make coding skill obsolete, or to allow `skids' and `n00bs' to act at high above their real level of knowledge, would appear to pose a direct threat to the core values of the hacker subculture.

 Indeed, members do show indications of trying to preserve traditional hacker practices, values, and habitus as they try to domesticate GenAI technologies within their social context \cite{silverstone_listening_1991}. Discussions clearly distinguish between low-and high-skill approaches to using generative AI-tools, with a focus on the importance of existing technical skill, creativity, and experimentation. Thus, there are clear attempts to preserve the existing systems of cultural capital and reframe these new technologies in the within the existing values of the hacker subculture -- i.e. we observe a process in which users are currently attempting to fit these technologies not only into their business models, but into the moral economy of their communities \cite{silverstone_listening_1991}.

In terms of disrupting the process of learning the skills and practices of cybercrime, we find that for many, the \textit{social} dimension of learning remains central. Members still want the community and interaction with others, and participation in a subculture, while shunning the idea of talking with an AI. Where we do see large, free-wheeling discussions of (for example, Dark AI services) these are not particularly practical, instead, reflecting a shared social experience of experimenting, speculating about, and enjoying the technology (especially where they are able to replicate existing products of hacker culture). Forum posts around Dark AI products appear therefore mostly to be a form of 
a shared subcultural practice rather than a serious disruption to the cybercrime ecosystem. These kinds of practices are often observed within deviant subcultures as a way of establishing, participating in, and performing shared identities.

In cybercrime communities, this kind of acting-out of `hacker' behaviour by neophytes has long been observed on forums and online cybercrime discussion boards. This (as with `Dark AI') acts as a shared celebration of cultural interest in 
cybercrime topics -- such as spending money on trivial criminal services, kit and hacking tools or guides, installing Kali Linux on the desktop, or accessing fake `red room' websites on the dark web -- while they are important in binding the community together and establishing norms, these not particularly relevant to the practical business of cybercrime as a crime phenomenon.

\subsection{Methodological thoughts}

The qualitative findings of the paper combined three sampling approaches, across LLM classification, topic modelling, and stratified random sampling. While LLM brought higher quality posts, these were highly biased towards longer vibe-coding posts. Topic modelling and random sample gave lower quality posts but a much more representative sample of what forum members are actually doing. Additionally while approximately 80\% of the positively labelled (non-other) posts from the LLM model were relevant, the precise classification was almost always wrong. This project found the use of local open-weight models for LLM-based analysis to not be ready for use on its own, and the random sampling was crucial in giving a better perspective.

\subsection{AI technologies more broadly}
The core assumptions and approaches used in the development of generative AI tools have been criticised long before the release of ChatGPT, highlighting the risks of high environmental and financial costs along with ethical risks such as entrenching status quo, majority, and discriminatory perspectives; promoting average rather than best practice; and reproducing abusive materials~\cite{bender_dangers_2021}.
The risks of anthropomorphising LLMs and believing that text that appears fluent implies intelligence or understanding and the risk that the ability to quickly generate plausible text can amplify hateful messages.
These risks have subsequently been deliberately ignored in the creation of generative AI tools~\cite{gebru_tescreal_2024}.
We see little evidence of discussion of these higher level risks within our data, however the keywords used were not selected to seek this out.

Studies have found that Generative AI tools have not always stood up to their hype. One study found that in early 2025 while software developers thought that using AI tools would improve their productivity, it actually increased time taken to complete tasks by 19\%~\cite{becker_measuring_2025}. Another found that using AI tools increased vulnerability and decreased reliability and maintainability of the software~\cite{fawzy_vibe_2025}. This aligns with the experiences we see reported in the cybercrime forums.

The cybercrime forums support a learning community and so as research into the use of AI in education has found `the risks of utilizing AI in education overshadow its benefits'~\cite{burns_new_2026}. We see the same conclusion arising among cybercrime forum users.

\subsection{Stand-alone complex or vibercrime?}

In answering our initial questions of maximal and minimal cases for adoption -- whether early adoption appears to be the start of a trajectory towards a Stand Alone Complex or Vibercrime model -- we find elements of both, though in far more prosaic terms. For the latter, it appears that, much as in the legitimate tech industry, coding assistants in particular are simply becoming integrated as productivity enhancers (namely, that the limited coding involved in cybercrime is following a similar trajectory to the legitimate industry). The desire of parts of the community to incorporate their core values of skill, experimentation and creativity into the adoption of these technologies suggests interesting potential evolutions in core ideas of what sophistication `is' in cybercrime (or rather, what ``L33tness'' and ``skiddiness'' look like for vibercrime). The use of LLMs in obfuscation of patterns in particular appears to be an important capability, though one that clearly requires substantial human effort. As AI-assisted coding (``vibe coding'' and ``vibe engineering'') continues to be adopted widely as part of mainstream coding practices, we observe this becoming de facto practice within the cybercrime ecosystem as well. As in the legitimate ecosystem, however, we can expect to see large amounts of technical debt begin to accrue in the cybercrime projects that adopt these practices -- particularly within more productised and infrastructural elements where maintenance and long-term stability are the driving factors in project success.

There are glimmers of the Stand Alone Complex here too, which blend science fiction with quotidian money-making schemes. We can see established communities -- which predate the LLMs and chatbots -- attempting to battle the machine learning systems of the major platforms with their own attempts at automation and large-scale infrastructure, in the pursuit of eking individually small profits out of the material political economy~\cite{mackenzie_da_2024}
of the advertising and influence platforms. Although LLMs are only a stepwise evolution in these practices, as the major platforms change their business models and infrastructure in response to the AI wave, we may see significant changes here as well. A move away from the highly distributed display ad-based material political economy that has characterised the Internet of the past twenty years towards a far more centralised model based around chatbots would produce serious opportunities for these actors. In particular, their skills in cat-and-mouse machine learning games with the platforms -- and in the management of large infrastructures of bots, websites, and accounts positioned on the key infrastructures that are feeding the chatbots -- would appear to pose serious opportunities for harm and profit.

Thus, we see largely at present the incorporation of these novel technologies into \textit{existing} criminal innovation trajectories rather than the formation of distinctive new trajectories proper. If these new trajectories do take shape, it may well be outwith the traditional cybercrime subculture.

\subsubsection{Recommendations and future work}

Our primary recommendation for industry, policymakers, and law enforcement can be summarised as `don't panic'. While clear harms are evident, adoption of AI tools in the cybercrime ecosystem has been piecemeal and non-transformative to date. Thus, patterns of adoption for technical tasks in mainstream software industries cannot be simplistically imputed across to cybercrime businesses (as these are largely not reliant on technical skill). Significant economic barriers to scale can be achieved through tuning and guardrailing of models and we suggest that this continue to be a focus for industry. Tuning to increase friction at the low-level (i.e. making individual use harder) can also be designed to simultaneously encourage saturation dynamics at scale when these technologies are used in volume automation. These do not and cannot deter motivated attackers, and need to be employed along with further elements of security design ~\cite{shumailov_thoughts_2026}. It is possible that our lukewarm findings on new markets simply reflect (as Shumailov argues~\cite{shumailov_thoughts_2026}) the fact that most mainstream industry has not successfully managed to implement agentic tools for full automation. It likely that if we see genuine transformative adoption in commercial settings, new criminal markets will emerge at that point. However, technical fixes alone will not solve issues that are fundamentally a product of social forces. Not least, many of the emerging effects in cybercrime are in capacity uplifts which are not cybercrime specific -- improvements in basic coding practices. Here, major policy levers may be outside the realm of law enforcement and justice entirely -- dealing with the consequences of labour market disruption and regulatory responses to changes to the platform ecosystem as a whole.

A more urgent area of focus than the effects on traditional cybercrime is in the use of these tools for harassment, particularly visible in campaigns of both targeted and mass-scale misogynistic harassment and in campaigns targeting vulnerable communities.
As prior research suggests, this is overwhelmingly enabled by social media platforms and mainstream, rather than specialist tools. Our research has found that technical interventions such as guardrailing can indeed have real effects in mitigating the scaling of these harm forms. However, this leaves a serious set of targeted harms enabled by specialist providers who have the time, motivation, and skills to work around guardrails, retrain models or combine GenAI and other tools. For these, as for the wider social harms pertaining to technological innovation, there are no purely design solutions. Instead serious regulation of business models and incentives, especially of the large mainstream platforms and services, and especially where corporate norms are favouring providing potentially harmful new technologies to a mass consumer base with little to no safety testing, or with actively harmful and misogynistic use cases being tacitly encouraged.

Future work in this area will undoubtedly trace these initial trajectories further as they develop. We would additionally encourage both qualitative interview studies with members of these communities, as well as a broadened focus that includes data from social media and actors outside the traditional cybercrime social scene. It may well be that the real innovation is happening in rather different communities entirely -- such as those centred around AI experimentation more fully.

\section{Limitations}
Data was sampled from underground cybercrime forums, which are open to anyone to join. This can provide a limited view across the wider cybercrime ecosystem, as it does not provide insight into invite-only, closed group, or larger threat actors. Furthermore, the dataset shows members who talk publicly to each other and form communities. There may be other members who only communicate privately, and these other types of members may behave differently.

This research is heavily dependent on our interpretation of the content, reading of forum posts, immersion in these communities' cultures, and computational analysis. Future research could usefully complement this with interview data from participants in these subcultures. Furthermore, even though people using use Gen AI or finding this useful, it does not necessarily mean it is changing the phenomenon in a meaningful way, or a way that would mimic wider changes in industry.

Our analysis covers up to mid 2026. It is entirely possible that some members of the cybercrime underground have found genuine utility in these tools for more technical matters. It is also possible that in future years a major novel innovation will escape its niche and trigger landscape change. However, we are confident in asserting that up to mid 2026, there is no empirical evidence of major \textit{structural} disruption of the `hacking' side of the cybercrime ecosystem by GenAI tools, despite a small number of widely-reported cases where these tools have been used. The automated and rapid production of 'boring' administrative and logistics tasks - such as convincing and detailed approach and phishing emails and conversations, visual and audio materials, and other forms of automation such as server administration or bot farming has been a far larger feature.

People may well find things and not talk about them on the forums. But -- if they had, we would typically expect to see some sort of secondary market for tools, adaptations or services emerge on the forums. It may be that these types of cybercrime communities have moved over to Discord, Telegram, and YouTube, and the forums are of decreasing relevance. Even so, forums are still some of the big active markets and communities, and we would expect to see more horizontal diffusion of capabilities here if there was real disruption in the other communities.

\section{Conclusion}\label{sec:conclusion}

In this paper, we have analysed early empirical data from a variety of large-scale digital sources from the cybercrime underground, and find that the reality is prosaic. Building on previous initial findings in the literature \cite{burton_ai_2025}, we find that AI is seeing some early adoption in existing large-scale, low-profit passive income schemes, in automating and scaling logistical aspects of cybercrime, and trivial forms of click fraud but is not giving rise to widespread disruption in the economy of technical skill. It is also not being widely used to reduce the skill barrier to entry for cybercrime-specific coding domains (which already rely heavily on old, pre-made resources, scripts and exploits). Instead, up until end 2025, it was replacing existing means of code pasting, error checking, and cheatsheet consultation, mostly for generic aspects of software development involved in cybercrime -- and largely for already skilled actors, with low-skill actors finding little utility in vibe coding tools compared to pre-made scripts. From 2026 onwards, vibe coding has been widely adopted in the underground, but again only effectively by already-skilled players. The role of jailbroken LLMs as hacking instructors is also overstated, given the prominence of deviant association and social learning in initiation -- new users value the social connections, subculture, and community identity involved in learning hacking and cybercrime skills as much as the knowledge itself.

In some ways this all suggests a more mature response from the cybercrime ecosystem than we see in industry (possibly because of the lack of a CEO and venture capital class). Many of the broader economic effects reflect patterns also observed in legitimate content and tech economies -- for saturation of overproduced low-quality content in particular. Therefore, we conclude that generative AI tools are having disruptive effects on cybercrime, but mostly to the economics and political economy of major platform infrastructure and the licit and illicit businesses that cluster around it, rather than to cybercrime directly. Finally, we note that in recent months anxiety over labour market disruption from these tools is increasing precipitously. This may end up being the most important way in which generative AI tools disrupt the cybercrime ecosystem -- mass layoffs, economic downturn and a cool job market pushing legitimate, more skilled developers into the underground communities of get rich quick schemes, fraud, and cybercrime.

\begin{appendices}\label{sec:appendix}

\section{Competing interests}
No competing interest is declared.

\section{Author contributions statement}
J.H., B.C. and D.R.T. devised the overall study design. J.H. designed the study methods, and processed data for quantitative and qualitative analysis. J.H. and B.C. analysed the results. B.C. led the write up of the manuscript, including background and results. J.H., B.C. and D.R.T. contributed to writing and reviewing of the manuscript.

\section{Acknowledgments}
Jack Hughes is supported by the European Research Council (ERC) under the European Union’s Horizon 2020 research and innovation programme (grant agreement No 949127).

\end{appendices}

\bibliographystyle{unsrt}
\bibliography{references}

\begin{biography}{{\includegraphics[width=77pt]{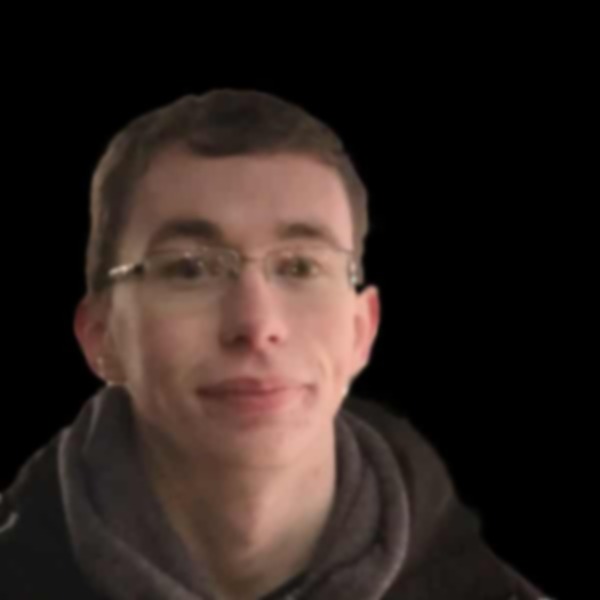}}}{\author{Jack Hughes.} is a Postdoc in the Cambridge Cybercrime Centre, Department of Computer Science and Technology at the University of Cambridge. As a Postdoc, he is exploring new research areas, around security, data, real-world systems, and the societal impacts of these. Previously during PhD studies, his research focused on developing novel tools and methodology to enable interdisciplinary research with large, messy datasets.}
\end{biography}

\begin{biography}{{\includegraphics[width=77pt]{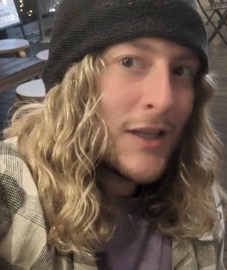}}}{\author{Ben Collier.} is Senior Lecturer in Science, Technology and Innovation Studies at the University of Edinburgh. Their research focuses on how digital infrastructures become sites for crime, harm, and security. Dr Collier has a mixed-methods, interdisciplinary background which draws on concepts and methods from both criminology and technology studies.}
\end{biography}

\begin{biography}{{\includegraphics[width=77pt]{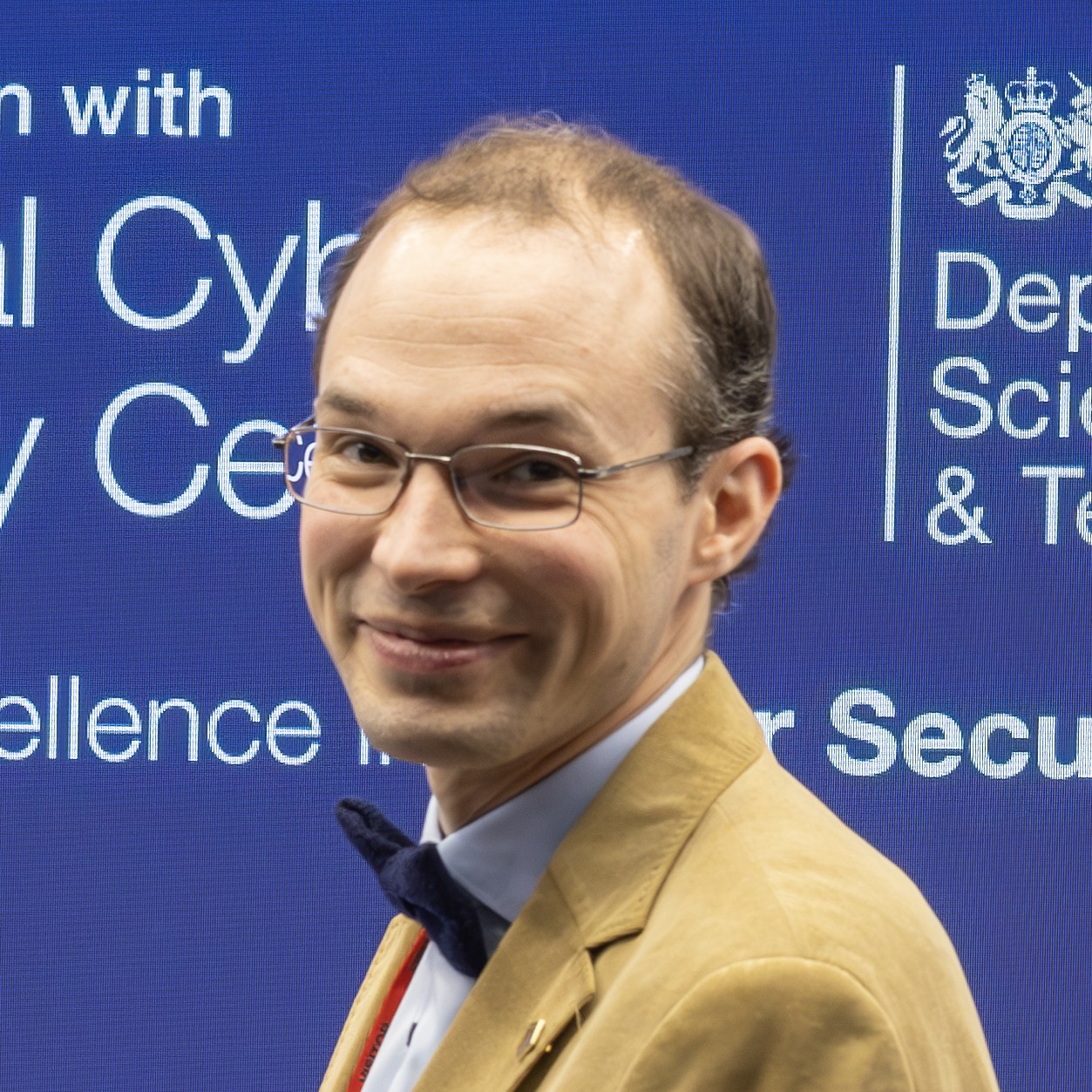}}}{\author{Daniel R. Thomas.} Is Director of the Academic Centre of Excellence in Cyber Security Research at the University of Strathclyde where his research focuses on measuring and understanding cybersecurity, cybercrime, and cyberresilience ethically. This enables monitoring of imprrovements, evaluating interventions, and informing regulators and helps provide better economic incentives. Dr Thomas has a technical background but often works across disciplines to understand the full picture.}
\end{biography}

\end{document}